\documentclass[12pt]{article}

\pdfoutput=1
\usepackage{slashed}
\usepackage{color, verbatim}
\usepackage{latexsym}
\usepackage{epsfig}
\usepackage{amsmath,accents}

\usepackage{amssymb}
\usepackage{graphicx}
\usepackage{bm}

\usepackage[font={small}]{caption}

\usepackage{cite}
\usepackage{hyperref}

\newcommand{\hhref}[1]{\href{http://arxiv.org/abs/#1}{arXiv:#1}}

\definecolor{myred}{rgb}{0.7, 0, 0}
\definecolor{myblue}{rgb}{0, 0, 0.7}
\definecolor{mygreen}{rgb}{0.04, 0.7, 0.5}
\hypersetup{colorlinks,citecolor=myred,linkcolor=myblue,urlcolor=myblue,linktocpage=true}

\makeatletter

\@addtoreset{equation}{section}
\makeatother

\newcommand{\be}{\begin{equation}}
\newcommand{\ee}{\end{equation}}
\newcommand{\bea}{\begin{eqnarray}}
\newcommand{\eea}{\end{eqnarray}}

\newcommand{\diag}{\operatorname{diag}}

\begin{document}

\thispagestyle{empty}

\begin{center}


\begin{center}

\vspace{.5cm}

{\Large\sc $g_\mu-2$ from Vector-Like Leptons in Warped Space}

\end{center}

\vspace{1.cm}

\textbf{
Eugenio Meg\'ias$^{\,a\,,b}$,  Mariano Quir\'os$^{\,c\,,d}$, Lindber Salas$^{\,c}$
}\\

\vspace{1.cm}
${}^a\!\!$ {\em {Max-Planck-Institut f\"ur Physik (Werner-Heisenberg-Institut), \\ F\"ohringer Ring 6, D-80805, Munich, Germany}}

\vspace{.1cm}
${}^b\!\!$ {\em {Departamento de F\'{\i}sica Te\'orica, Universidad del Pa\'{\i}s Vasco UPV/EHU, \\ Apartado 644,  48080 Bilbao, Spain}}

\vspace{.1cm}
${}^c\!\!$ {\em {Institut de F\'{\i}sica d'Altes Energies (IFAE),\\ The Barcelona Institute of  Science and Technology (BIST),\\ Campus UAB, 08193 Bellaterra (Barcelona) Spain}}

\vspace{.1cm}
${}^d\!\!$ {\em {ICREA, Pg. Llu\'is Companys 23, 08010 Barcelona, Spain}}

\end{center}

\vspace{0.8cm}

\centerline{\bf Abstract}
\vspace{2 mm}
\begin{quote}\small
The experimental value of the anomalous magnetic moment of the muon, as well as the LHCb anomalies, point towards new physics coupled non-universally to muons and electrons. Working in extra dimensional theories, which solve the electroweak hierarchy problem with a warped metric, strongly deformed with respect to the AdS$_5$ geometry at the infra-red brane, the LHCb anomalies can be solved by imposing that the bottom and the muon have a sizable amount of compositeness, while the electron is mainly elementary. Using this set-up as starting point we have proven that extra physics has to be introduced to describe the anomalous magnetic moment of the muon. We have proven that this job is done by a set of vector-like leptons, mixed with the physical muon through Yukawa interactions, and with a high degree of compositeness. The theory is consistent with all electroweak indirect, direct and theoretical constraints, the most sensitive ones being the modification of the $Z\bar\mu\mu$ coupling, oblique observables and constraints on the stability of the electroweak minimum. They impose lower bounds on the compositeness ($c\lesssim 0.37$) and on the mass of the lightest vector-like lepton ($\gtrsim 270$ GeV). Vector-like leptons could be easily produced in Drell-Yan processes at the LHC and detected at $\sqrt{s}=13$ TeV.

\end{quote}

\vfill

\newpage

\tableofcontents

\newpage
\section{Introduction}
\label{introduction}

In spite of the fact that, so far, LHC has found no direct evidence of new physics (NP) beyond the Standard Model (BSM),  there are several hints of lepton flavor universality NP. Two of them are related to the muon lepton flavor, as the $B\to K^*\mu^+\mu^-$ anomalies~\cite{Aaij:2014ora,Aaij:2015oid} and the anomalous magnetic moment (AMM) of the muon~\cite{Bennett:2006fi}, so one could suspect that both of them could be related to the same kind of NP. 

Moreover, as the Standard Model (SM) has a naturalness problem (the so-called hierarchy problem) it would be rewarding to accommodate the solutions to present (or future) experimental anomalies within theories solving the hierarchy problem. The most popular solutions to the hierarchy problem are provided by supersymmetric theories (where the electroweak scale is protected by  supersymmetry) and by theories with a warped extra dimension~\cite{Randall:1999ee} (where the electroweak scale is provided by the Planck scale after warping along the extra dimension)~\footnote{In particular theories with a warped extra dimension are dual to theories with a strongly coupled sector and a composite Higgs, which are by themselves theories solving the hierarchy problem as, at the compositeness scale, the Higgs melts into its components.}. 

We will concentrate on the latter class of theories, i.e.~in theories with a warped extra dimension. In particular we will consider theories with two branes, an ultra-violet (UV) and an infra-red (IR) brane, and a stabilizing field $\phi$ strongly deforming the AdS$_5$ metric near the IR-brane. This strong deformation makes it possible to accommodate the SM in the bulk, without an additional custodial gauge symmetry, consistently with all electroweak and flavor constraints, thanks to a naked metric singularity in the extra dimension (soft-wall metric) outside the physical interval~\cite{Cabrer:2009we, Cabrer:2010si, Cabrer:2011fb, Cabrer:2011vu, Cabrer:2011mw, Carmona:2011ib, Cabrer:2011qb, deBlas:2012qf,Quiros:2013yaa, Megias:2015ory}. We have recently shown that in this theory one can easily accommodate the LHCb anomalies provided that the left-handed muon (and bottom quark)  has some degree of compositeness~\cite{Megias:2016bde}.

In this paper we will consider the other muon anomaly: the AMM of the muon. At the tree level the muon predicts a magnetic moment $\vec M_\mu=g_\mu \frac{e}{2m_\mu}\vec S_\mu$ with gyromagnetic ratio $g_\mu=2$. Loop effects predict a deviation with respect to the tree level value which is parameterized by the ratio (AMM)
\be
a_\mu=\frac{g_\mu-2}{2}\,.
\ee
The SM gives a very precise prediction of the AMM of the muon~\cite{Agashe:2016kda}. In particular a recent update of the hadronic vacuum polarization contribution to the AMM~\cite{Davier:2016iru} yields a value $a_\mu^{\rm SM}$ which deviates with respect to the experimental determination $a_\mu^{\rm exp}$~\cite{Bennett:2006fi} by $\sim 3.6\, \sigma$, i.e.
\be
\Delta a_\mu\equiv a_\mu^{\rm exp}-a_\mu^{\rm SM}=(2.74\pm 0.76)\times 10^{-9}
\label{deviation}
\ee

There are a number of proposals aiming to explain the experimental value of the AMM of the muon by means of new BSM physics. Many of these proposal invoke physics unrelated to the solution of the hierarchy problem, as introducing $Z'$ gauge bosons, extra fermions, scalars, vectors, or lepto-quarks. For some recent papers see Refs.~\cite{Biggio:2016wyy,Altmannshofer:2016oaq,ColuccioLeskow:2016dox,Batell:2016ove,Freitas:2014pua,Altmannshofer:2016brv} and references therein. There are also a number of explanations of the AMM of the muon in the context of supersymmetric theories, which essentially select the space of supersymmetric parameters such that there can be enhanced contributions to $\Delta a_\mu$. For a review of supersymmetric contributions to $\Delta a_\mu$ see Ref.~\cite{Stockinger:2006zn}.

In this work we will consider a possible explanation of the AMM of the muon, in the context of theories solving the hierarchy problem by means of a warped extra dimension. We will do that in soft-wall metric models, where the SM fields can propagate in the bulk of the extra dimensions without invoking an extra gauge custodial symmetry, as described in Refs.~\cite{Cabrer:2009we, Cabrer:2010si, Cabrer:2011fb, Cabrer:2011vu, Cabrer:2011mw, Carmona:2011ib, Cabrer:2011qb, deBlas:2012qf,Quiros:2013yaa, Megias:2015ory}. As these theories can accommodate the fermion flavor problem of the SM, by means of particular values of the parameters localizing the fermions along the extra dimension, we will adopt the particular configuration of these parameters which provide a natural solution to the LHCb anomaly, as done in Ref.~\cite{Megias:2016bde}. This configuration will settle a starting point for analyzing the AMM of the muon. The outline of the rest of this paper will be as follows.

In Sec.~\ref{sec:model} we introduce the model of warped extra dimension we will be using throughout this paper. We show the consistency of the model with the main electroweak constraints, in particular the oblique observables and the $Z\bar\mu\mu$ coupling. We also show how the model can accommodate the LHCb anomalies, which motivates the choice of the localizing (compositeness) parameters in the muon sector which will be used in the rest of the paper. Finally we show that the minimal version of the model is unable to explain the AMM of the muon, which motivates the introduction of vector-like leptons (VLL) with a Yukawa mixing to the muon sector. The formalism of VLL propagating in the bulk of the extra dimension, and their boundary conditions, is covered in Sec.~\ref{sec:VLL}. We show that masses $\gtrsim 1$ TeV imply fermions localized toward the IR brane, i.e.~fermions with a certain degree of compositeness in the dual theory. The gauge interactions of VLL with the gauge boson $Z$ and the Kaluza-Klein (KK)-modes $Z_n,\,\gamma_n$ are studied in Sec.~\ref{sec:gauge}. In particular the couplings of VLL with the KK-modes are very strong in the deep IR (for VLL localized toward the IR brane) while they are very week for VLL localized toward the UV brane. The former behavior will partly determine the posterior explanation of the AMM of the muon. The mixing of VLL with the muon through Yukawa interactions will be studied in Sec.~\ref{sec:Yukawa}. In particular the physical mass eigenstates will be found by diagonalization of the mixed VLL-muon mass matrix, through some unitary matrices $U_{L,R}$, providing some mixing angles between the sector of VLL and that of the muon. The gauge couplings studied in Sec.~\ref{sec:gauge} will be then modified by the presence of the mixing in the matrices $U_{L,R}$. As the mixing between VLL and the muon sector must be small as implied by electroweak constraints, the corresponding entries in the matrices $U_{L,R}$ must be small which allows an explicit analytical approximation for $U_{L,R}$ as performed in Sec.~\ref{sec:analytical}. This analytical approximation will simplify all couplings and will allow a much simpler treatment and understanding of further calculations in this paper. Moreover as electroweak constraints in the muon sector are very strong the accuracy of our analytical approximation will show up to be an extremely efficient one. A further simplification (this time a purely instrumental one) will be done in Sec.~\ref{sec:cLcR}, where we will impose a simplifying assumption: the localization parameters of doublet ($c_L$) and singlet ($c_R$) VLL are equal ($c_{L}=c_R\equiv c$). This assumption reduces the number of free parameters and allows a simplification of the matrices $U_{L,R}$. Using this particular case we will study the five-dimensional (5D) Yukawa couplings and found to lie in the perturbative region. In Sec.~\ref{sec:Zmumu} we single out the strongest electroweak constraint: the $Z\bar\mu\mu$ coupling, which gets modified by the mixing of the muon with VLL. We have proven that it constrains the absolute value of the off-diagonal elements of the unitary matrices $U_L^{31}$ and $U_R^{21}$ to be $\lesssim 0.02$, which justifies \textit{a posteriori} the approximation done in Sec.~\ref{sec:analytical}. Using the previous constraints we have computed in Sec.~\ref{sec:Deltaamu} the contribution of VLL and the vectors $Z,\,W,\,Z_n,\,\gamma_n,\,W_n$, and the Higgs $H$, fields to the AMM of the muon. We have shown the region in the parameter space where the value of the AMM can be in agreement with the experimental result of Eq.~(\ref{deviation}). In particular we have proven that the agreement implies that VLL have a high degree of compositeness, i.e.~that they are localized toward the IR brane (in particular that $c\lesssim 0.42$). The rest of constraints (except for the $Z\bar\mu\mu$ constraint) are analyzed in Sec.~\ref{sec:other}. We study constraints from oblique observables, from LHC data on the $H\to\gamma\gamma$ decay when VLL run inside the loop, from the stability of the electroweak minimum as VLL accelerate the running of the Higgs quartic coupling towards negative values, and finally from collider phenomenology as the VLL can be pair produced by Drell-Yan processes at hadron colliders. All these constraints reduce the size of the region allowed by VLL and leave a permitted region where $c\lesssim 0.37$ and the mass of VLL is $\gtrsim 270$ GeV. Finally our conclusions, and some comments about possible extensions of this work, are drawn in Sec.~\ref{sec:conclusions}.

\section{The model}
\label{sec:model}

We will review in this section the main aspects of the 5D warped model proposed and developed in Refs.~\cite{Cabrer:2009we, Cabrer:2010si, Cabrer:2011fb, Cabrer:2011vu, Cabrer:2011mw, Carmona:2011ib, Cabrer:2011qb, deBlas:2012qf,Quiros:2013yaa, Megias:2015ory}.
We assume the Higgs doublet to be a 5D field, so that
it propagates in the bulk. Splitting the degrees of freedom into Goldstone modes $\chi(x,y)$, vacuum expectation (background) value $h(y)$ and physical fluctuations $\xi(x,y)$ we can rewrite the Higgs field as
\be
H(x,y)=e^{i\chi(x,y)}\left( \begin{array}{c}  0\\ h(y)+\frac{1}{\sqrt{2}}\xi(x,y) \end{array}   \right)\,.
\ee
Electroweak symmetry breaking (EWSB) is triggered by an IR brane potential, whereas additional mass terms are introduced for the Higgs in the bulk and at the UV brane. The full Higgs potential is then
\be
V(H)=M^2(\phi) |H|^2+M_0|H|^2\delta(y)+\left(-M_1|H|^2 +\gamma|H|^4 \right)\delta(y-y_1)\,,
\ee
with
\be
M^2(\phi)=\alpha k \left[\alpha k-\frac{2}{3}W(\phi)  \right]\,.
\ee
where $\phi$ is the 5D bulk propagating field which stabilizes the size of the extra dimension at the value $y=y_1$,  $k$ a parameter with mass dimension related 
to the curvature along the fifth dimension~\cite{Cabrer:2009we}, and $W(\phi)$ the superpotential which fixes the gravitational background metric $A(y)$ such that
\be
ds^2=e^{-2A(y)}\eta_{\mu\nu}dx^\mu dx^\nu+dy^2
\ee

The dimensionless parameter $\alpha$ controls the localization of the Higgs wavefunction and can thus be
connected to the amount of tuning related to the hierarchy problem~\footnote{In fact solving the whole hierarchy problem amounts to fixing $A(y_1)\simeq 35$.}.
The Higgs background $h(y)$ has the required exponential shape
\be
h(y) = h_0 e^{\alpha k y}
\label{higgsprofile}
\ee
and it can be easily checked that
the fine-tuning is avoided for large enough values of $\alpha$, i.e.
\be
\alpha \gtrsim \alpha_1 = \frac{2 A_1}{ky_1}\,,
\ee
where $A_1\equiv A(y_1)$, which correspond to localizing the Higgs background profile towards the IR brane.

The SM fermions are realized in our scenario as chiral zero modes of 5D fermions.
The localization of the different fermions is determined by their 5D (Dirac) mass term.
The mass term for the 5D fermions
can be conveniently chosen as $M_{f_{L,R}}(y)=\mp c_{f_{L,R}} W(\phi)/6$
where the upper (lower) sign applies for fields with left-handed (right-handed) zero modes\cite{Cabrer:2011qb}.

In this paper we will primarily focus in the leptonic sector and, in particular, in the second generation of leptons. We will introduce the notation for the 5D leptons as
\be
\begin{pmatrix}\nu_i\\ \ell_i\end{pmatrix},\quad E_i
\ee
where $i$ is a generation index, the doublets have hypercharge $Y=-1/2$ and the singlets hypercharge $Y=-1$. In the following 
we will  just consider the second lepton generation ($i=2$) and will drop the generation index. We will impose boundary conditions such that the zero mode of $\ell$ has only left-handed chirality $\ell_L$ and the zero mode of $E$ only right-handed chirality $E_R$, where the contribution from the zero-modes is
\be
\ell_L(x,y)_L=\ell_L(y)\mu_L(x)+\cdots,\quad E_R(x,y)=E_R(y)\mu_R(x)+\cdots
\label{muones}
\ee
and the ellipses indicate the contribution from the non-zero KK-modes The 5D wave functions for the zero modes are given by
\be
\ell_L(y)=\frac{e^{(2-c_{\mu_L})A(y)}}{\displaystyle\left(\int dy\, e^{A(1-2 c_{\mu_L})} \right)^{1/2}},\quad E_R(y)=\frac{e^{(2-c_{\mu_R})A(y)}}{\displaystyle\left(\int dy\, e^{A(1-2 c_{\mu_R})} \right)^{1/2}}
\label{muon}
\ee
where $c_{\mu_L}$ provides the 5D Dirac mass of the doublet and $c_{\mu_R}$ that of the singlet.

In this paper we will use the superpotential formalism~\cite{DeWolfe:1999cp} and consider  the 5D gravitational background $A(y)$ determined by the superpotential~\cite{Megias:2015ory}
\be
W(\phi)=6k\left(1+e^{a_0 \phi}  \right)^{b_0}
\ee
where $a_0$ and $b_0$ are real dimensionless parameters. This model has been analyzed thoroughly for different values of the superpotential parameters 
in Refs.~\cite{Megias:2015ory,Megias:2015qqh}. The main feature of this kind of gravitational (soft-wall) models is that the 5D metric has a naked singularity~\cite{Cabrer:2009we} outside (but near) the physical interval and their
prediction for electroweak observables is greatly suppressed with respect to that of the AdS$_5$ case~\cite{Cabrer:2010si}, as we will now review.

\subsection{Oblique corrections from KK-modes}
The $S$ and $T$ parameters, contributing to oblique electroweak observables, are given by the general expressions~\cite{Peskin:1991sw}
\be
S=-16\pi \Pi^\prime_{3Y}(0),\quad T=\frac{4\pi}{s_W^2 c_W^2}\left[\Pi_{11}(0)-\Pi_{33}(0) \right]\,.
\label{ST}
\ee
%
%
Their contribution from the gauge KK modes was already considered in Refs.~\cite{Megias:2015ory}. They are given by the following expressions~\cite{Cabrer:2011fb}
\begin{align}
\alpha_{EM} \Delta T & =s^2_W \frac{m_Z^2}{\rho^2}k^2 y_1\int_{0}^{y_1}
\left[1-\Omega_h(y)\right]^2e^{2A(y)-2A_1} dy\,,
\nonumber\\
\alpha_{EM} \Delta S & =8c^2_Ws^2_W \frac{m_Z^2}{\rho^2}k^2 y_1\int_{0}^{y_1}
\left(1-\frac{y}{y_1}\right)\left[1-\Omega_h(y)\right]e^{2A(y)-2A_1} dy\,,\nonumber\\
\alpha_{EM}\Delta U & \simeq 0\,,
\end{align}
where $\rho\equiv ke^{-A(y_1)}$ and
\be
\Omega_h(y)=\frac{\omega(y)}{\omega(y_1)},\quad \omega(y)=\int_{0}^{y}
h^2(\bar y)e^{-2A(\bar y)} d\bar y \,.
\ee
These expressions include the leading contributions, which are due to the tree-level mixing of the
SM gauge bosons with the massive vector KK modes. 

\begin{figure}[htb]
\centering
\includegraphics[width=8.5cm]{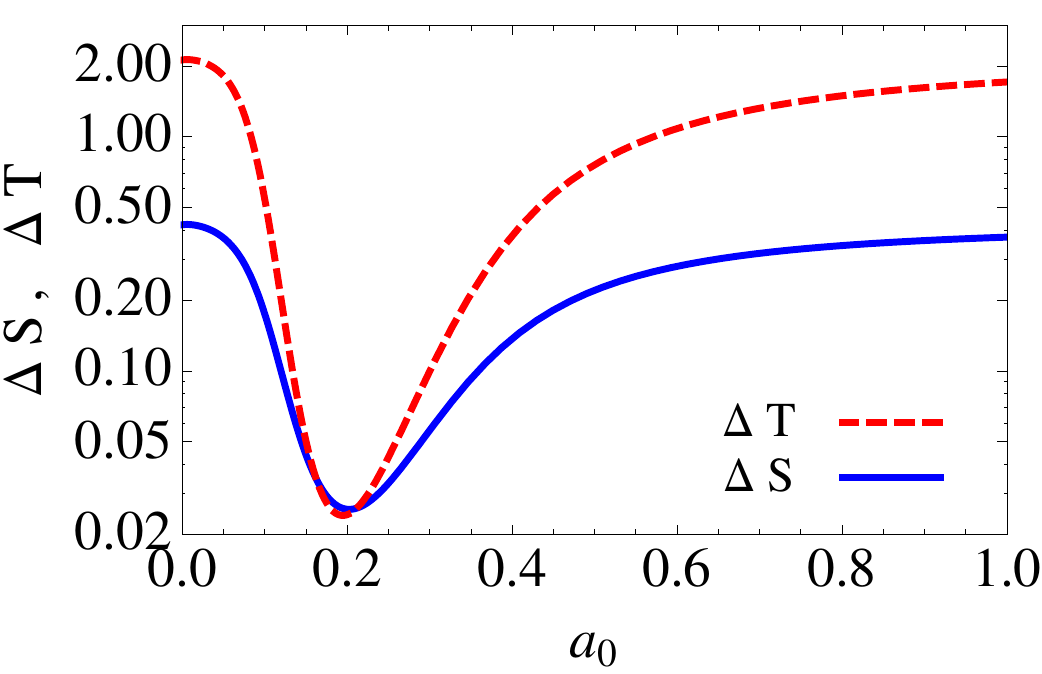} \hfill
\caption{\it Contribution to the $S$ and $T$ parameters from the gauge KK modes as a function of~$a_0$. We have considered  $b_0=1.5$ and $m_{KK}=2$~TeV.}
\label{fig:STa0}
\end{figure}

We show in Fig.~\ref{fig:STa0} the KK contribution to the oblique parameters, for $m_{KK}=2$ TeV (where $m_{KK}$ is the mass of the first KK mode of gauge bosons in the absence of electroweak breaking)  and $b_0=1.5$, as a function of~$a_0$~\footnote{For other values of $m_{KK}$ and the parameter $b_0$ see Ref.~\cite{Megias:2015ory}.}. In particular we can see from Fig.~\ref{fig:STa0} that for values $a_0\simeq 0.2$, their contribution is tiny: $\Delta S\simeq 0.0257$, $\Delta T\simeq 0.0244$. Therefore this small contribution, and possibly other kind of new physics contributing to the parameters $\Delta S$ and $\Delta T$, as we will see in the next section, leaves room to accommodate the experimental values~\cite{Agashe:2016kda}:
\be
S=0.07\pm 0.08,\quad T=0.1\pm 0.07,\quad  \textrm{(91\% correlation)}\,.
\label{eq:ST}
\ee

In this paper we will then consider, from now on, the particular
set of `gravitational' parameters given by: 
\be
a_0=0.2,\ b_0=1.5,\ \alpha=\alpha_1,\ A_1=35,\  m_{KK}=2 \textrm{ TeV.}
\label{set}
\ee

\subsection{$\delta g_{Z\overline{\mu}\mu}$ from KK-modes} 
\label{subsec:Zmumu}

As the new physics considered in this paper concerns the muon sector, an obvious strong effect is on modifications of the coupling of the $Z$ gauge boson with the physical muon. 
%
The main correction in this theory to $\delta g_{\mu_{L,R}}/g_{\mu_{L,R}}$ comes from the mixing of the $Z$ gauge boson with its KK-modes and from the mixing 
of the muon zero mode with its KK modes.  The resulting effect can be written as~\cite{Cabrer:2011qb}
\begin{equation}\label{eq:delta_g}
\delta g_{\mu_{L,R}}= - g_{\mu_{L,R}}^{SM}m_Z^2\widehat \alpha_{\mu_{L,R}}\pm \frac{g}{c_W} \frac{v^2}{2} \widehat\beta_{\mu_{L,R}}\,,
\end{equation}
where
\begin{align}
\widehat\alpha_{\mu_{L,R}}= & y_1\int_0^{y_1} e^{2A}\left(\Omega_h-\frac{y}{y_1}\right)\left(\Omega_{\mu_{L,R}}-1\right)dy\,,\nonumber\\
\widehat\beta_{\mu_{L,R}}= & Y_\mu^2\int_0^{y_1} e^{2A}\left( \frac{d \Omega_{\mu_{R,L}}}{dy}\right)^{-1}\left(\Gamma_\mu-\Omega_{\mu_{R,L}}\right)^2dy\,,\label{eq:defs1}
\end{align}
with $Y_\mu$ the muon Yukawa coupling and
\be\label{eq:defs2}
\Omega_{\mu_{L,R}}=\frac{\displaystyle \int_0^y e^{(1-2c_{\mu_{L,R}})A}dy}{\displaystyle \int_0^{y_1 }e^{(1-2c_{\mu_{L,R}})A}dy}\,,\qquad
\Gamma_\mu=\frac{\displaystyle \int_0^y he^{-(c_{\mu_L}+c_{\mu_R})A}dy}{\displaystyle \int_0^{y_1} he^{-(c_{\mu_L}+c_{\mu_R})A}dy}\,.
\ee
It is easy to recognize that the two terms in Eq.~(\ref{eq:delta_g}) correspond, respectively, to the effects of the massive vectors and of the fermion KK modes.

\begin{figure}[htb]
\centering
\includegraphics[width=9cm]{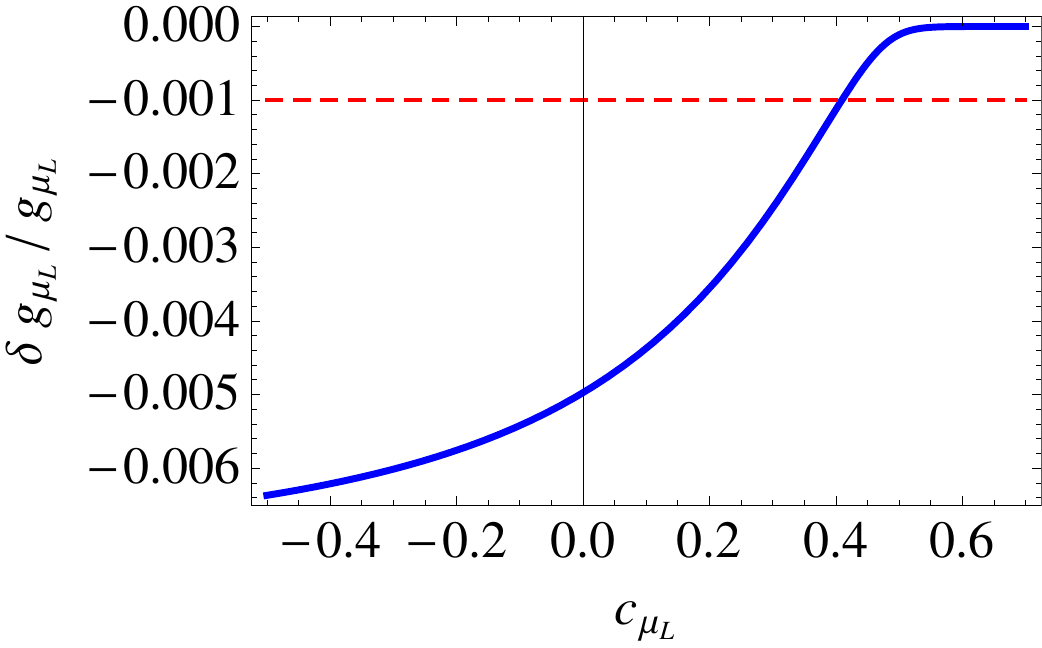} \hfill
\caption{\it Contribution to $\delta g_{\mu_L}/g_{\mu_L}$ from KK modes. The horizontal dashed line corresponds to $|\delta g_{\mu_L} / g_{\mu_L}| = 10^{-3}$. We have considered $c_{\mu_R}=0.5$.}
\label{fig:deltag}
\end{figure}

For the metric we are considering in this paper with $a_0=0.2$ and $b_0=1.5$, and for the KK gauge bosons with mass $m_{KK}=2$ TeV, the values we obtain for $\delta g_{\mu_{L,R}}/g_{\mu_{L,R}}$ are shown in the plot of Fig.~\ref{fig:deltag}.
We can see that for $c_{\mu_R}\gtrsim 0.5$, the experimental constraint $|\delta g_{\mu_{L,R}}/g_{\mu_{L,R}}| \lesssim  10^{-3}$~\cite{Agashe:2016kda} imposes $c_{\mu_L}\gtrsim 0.4$. In the rest of this paper we will fix $c_{\mu_L}=0.4$, a value consistent with the LHCb anomaly as we will see in the following.

\subsection{The $B\to K^* \mu^+\mu^-$ anomaly from KK-modes}

\begin{figure}[htb]
\centering
\includegraphics[width=8.5cm]{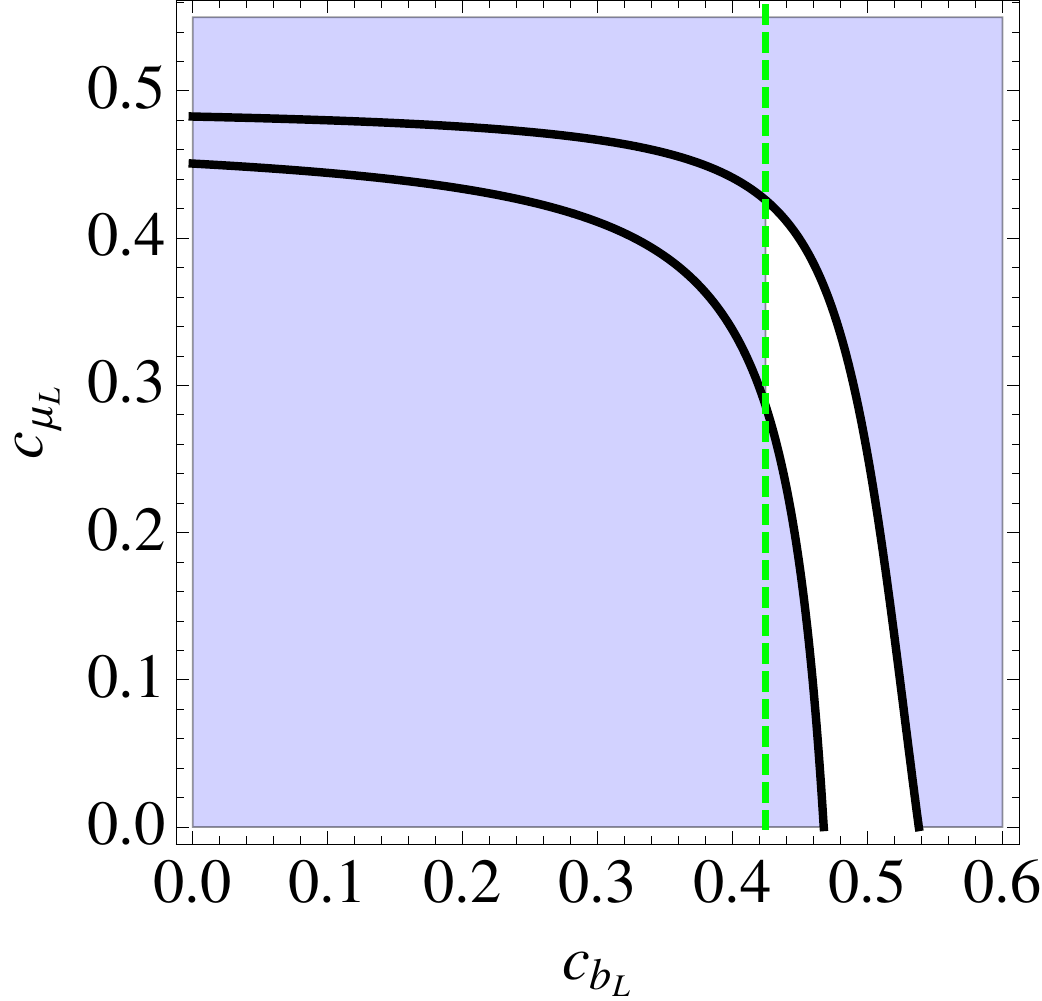} \hfill
\caption{\it Region in the plane $(c_{b_L},c_{\mu_L})$ that accommodates $\Delta C_9\in[-1.67,-0.39]$. We have also indicated the bound (vertical line) from flavor physics in the botton sector corresponding to $c_{b_L} > 0.424$ (see Refs.~\cite{Megias:2016bde,Megias:2016jcw} for further details). We have considered  $c_{\mu_R}=0.5$, $c_{s_L}=0.6$
.}
\label{fig:DC9}
\end{figure}

We have recently shown that this theory can naturally accommodate the LHCb anomaly if the muon has a certain degree of compositeness~\cite{Megias:2016bde,Megias:2016jcw}. In fact the contribution to the Wilson coefficient of the relevant $\Delta F=1$ operator $\mathcal O_9=(\bar s_L \gamma_\mu b_L)(\bar\mu\gamma^\mu \mu)$ can be written as
\be
\Delta C_9=-\sum_{X=Z,\gamma}\sum_n \frac{\sqrt{2}\pi g^{X_n}_{\mu_V}(g^{X_n}_{b_L}-g^{X_n}_{s_L})}{G_F\alpha M_n^2} 
\label{C9}
\ee
where the fitted values from experimental data are $\Delta C_9\in[-1.67,-0.39]$~\cite{Descotes-Genon:2015uva}. The couplings $g^{X_n}_{\mu_V}$, $g^{X_n}_{b_L}$ and $g^{X_n}_{s_L}$ are provided by the overlapping of the wave functions of the corresponding fermion and the KK-gauge bosons. For the calculation of $g^X_{b_L}$ we choose $c_{b_L}=0.44$ (a value passing all the constrains in Ref.~\cite{Megias:2016bde}). 
Moreover for the values that we will consider in the present paper, $c_{\mu_L}=0.4$ and $c_{\mu_R}=0.5$, we obtain a value $\Delta C_9= -0.464$ which is consistent with an explanation of the LHCb anomaly and passes all the precision tests from Ref.~\cite{Megias:2016bde}. We show in Fig.~\ref{fig:DC9} the parameter space region in the $(c_{b_L}, c_{\mu_L})$ plane that allows to fit the flavor anomalies. %

\subsection{$\Delta a_\mu$ from muon KK-modes}
The theory described in Sec.~\ref{sec:model} provides a framework where the (minimal) Standard Model propagates in the warped extra dimension thus solving the Higgs hierarchy problem, consistently with all electroweak precision data, and providing a solution to the quark and lepton flavor problem by fermion localization in the extra dimension, consistently with flavor data~\cite{Cabrer:2011qb}. Moreover, as was shown in the previous section and in Ref.~\cite{Megias:2016bde}, the theory could accommodate some of the recently observed flavor anomalies, in particular the  $B\to K^* \mu^+\mu^-$ anomaly. 

The other anomaly in the muon sector, as explained in Sec.~\ref{introduction}, is the experimental value of the muon anomalous magnetic moment $a_\mu$. The exchange of muon KK-modes along with $Z$ and $\gamma$ KK-modes, in diagrams similar to those of Fig.~\ref{fig:Delta-amu-Z} (with obvious modifications), should be good candidates to explain the experimental value required for $\Delta a_\mu$. However because of the structure of the 5D muon sector in Eqs.~(\ref{muones}) and (\ref{muon}) the chirality flip in the triangular diagram contributing to $\Delta a_\mu$ is suppressed by an $\mathcal O(m_\mu/m_{KK})$ factor leading to a too small effect unable to cope with the experimental result~\footnote{We thank Giuliano Panico for a discussion on this point.}. There are in the literature similar scenarios in Randall-Sundrum models that are unable to accommodate the experimental value of $\Delta a_\mu$ by at least one order of magnitude, see e.g.~\cite{Beneke:2012ie,Moch:2014ofa}.

As a consequence, the theory we are considering has to be enlarged to reproduce the experimental value of the muon anomalous magnetic moment. We will provide, in the rest of this paper, an extra sector, containing vector like leptons propagating in the bulk of the extra dimension, which mix with the muon sector through Yukawa interactions, providing the required sizable chirality flip in $\Delta a_\mu$. As we will see next, the required mixing is consistent with all present experimental and theoretical constraints in the very sensitive muon sector.

\section{Vector like leptons}
\label{sec:VLL}

We will now introduce vector-like leptons 
\be
D(x,y)=\begin{pmatrix}N(x,y) \\ L(x,y) \end{pmatrix} _{-1/2},\quad R(x,y)_{-1}
\label{defVLF}
\ee
transforming as a doublet and a singlet under $SU(2)_L$, respectively, and with the same hypercharge as the SM leptons. 
We will give them 5D Dirac masses $M_{L,R}(y)$ depending on the constants $c_L$ and $c_R$, and boundary 
conditions such that the zero modes are four-dimensional (4D) Dirac spinors with mass eigenvalues $M_{\mathcal L}(c_L)$ and $M_{\mathcal R}(c_R)$, 
respectively. 

In order to figure out what are the boundary conditions (BC) that we need to impose
to generate   $M_{\mathcal L,\mathcal R} \neq 0$, we  write the zero modes decomposition as
\begin{align}
N_{L,R}(x,y)&=N_{L,R}(y) \mathcal N_{L,R}(x),\quad
L_{L,R}(x,y)=L_{L,R}(y) \mathcal L_{L,R}(x)\nonumber\\
R_{L,R}(x,y)&=R_{L,R}(y) \mathcal R_{L,R}(x)
\end{align}
where the wave functions are normalized such that
\be
\int e^{-3A}L_{L,R}^2(y)dy=\int e^{-3A}R_{L,R}^2(y)dy=1\ .
\ee
and $N_{L,R}(y)\equiv L_{L,R}(y)$ from the $SU(2)_L$ invariance. Defining  the new functions
\be
\widehat{L}_{L,R}(y)=e^{-2A}L_{L,R}(y), \quad
\widehat{R}_{L,R}(y)=e^{-2A}R_{L,R}(y),
\label{}
\ee
the Dirac equations for  $\widehat{L}_{L,R}$ and $\widehat{R}_{L,R}$ are written as 
\begin{align}
M_{\mathcal L}e^{A}\widehat{L}_{R,L}(y) &=(M_{L}(y)\mp \partial_{y})\widehat{L}_{L,R}(y),\nonumber\\
M_{\mathcal R}e^{A}\widehat{R}_{R,L}(y) &=(M_{R}(y)\mp \partial_{y})\widehat{R}_{L,R}(y). \label{VLF1}
\end{align}

\begin{figure}[htb!]
\begin{center}
\includegraphics[scale=1]{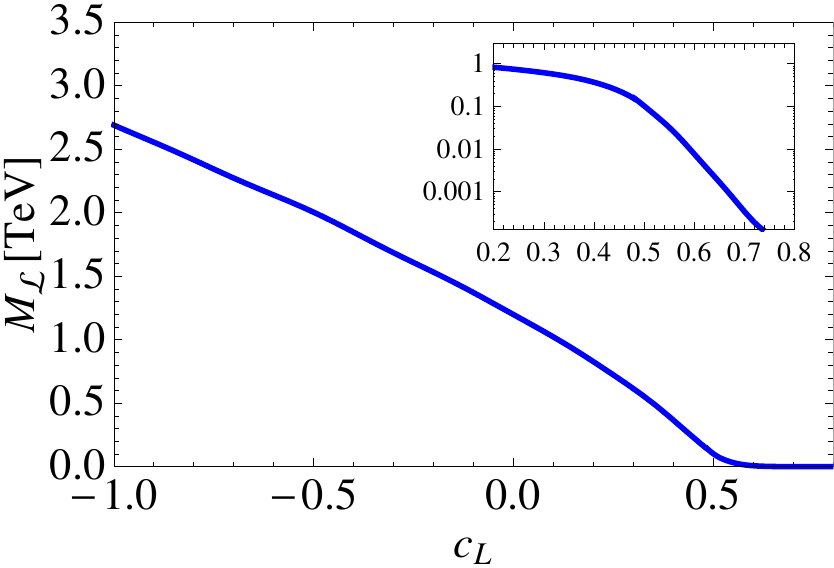}
\caption{\it VLL mass $M_{\mathcal{L}}$ as a function of the parameter $c_L$. The inserted figure corresponds to a logarithmic plot in the regime in which $M_{\mathcal{L}}$ is small. The same plot would apply of course for $M_{\mathcal{R}}$ as a function of $c_R$. }
\label{fig:ML}
\end{center}
\end{figure}

Imposing the BC as~\footnote{We thank O. Pujolas for discussions on this point.}
\begin{align}
(M_{L}+\partial_{y})\widehat{L}_{L}\lvert_{y=0}=0 ,\quad   \widehat{L}_{L}\lvert_{y=y_1}=0,\nonumber\\
\widehat{L}_{R}\lvert_{y=0}=0,  \quad  (M_{L}-\partial_{y})\widehat{L}_{R}\lvert_{y=y_1}=0, \label{VLF2}
\end{align}
and similarly
\begin{align}
(M_{R}+\partial_{y})\widehat{R}_{L}\lvert_{y=0}=0, \quad \widehat{R}_{L}\lvert_{y=y_1}=0 ,\nonumber\\
 \widehat{R}_{R}\lvert_{y=0}=0,\quad  (M_{R}-\partial_{y})\widehat{R}_{R}\lvert_{y=y_1}=0.\label{VLF3}
\end{align}
it is easy to see that $M_{\mathcal{L},\mathcal{R}} \neq 0$. The proof goes as follows: assuming  $M_{\mathcal{L}}=0$, then from (\ref{VLF1}) we would have the solution 
\begin{align}
\widehat{L}_{L,R}(y)= \widehat{n}_{L,R} e^{\pm \int^y M_{L}dy},
\end{align}
where  $\widehat{n}_{L,R}$ are constants determined by  (\ref{VLF2}). The BC  $(M_{L}+\partial_{y})\widehat{L}_{L}\lvert_{y=0}=0$ and
$(M_{L}-\partial_{y})\widehat{L}_{R}\lvert_{y=y_1}=0$ are  automatically satisfied by the Dirac equation (\ref{VLF1}), while the BC
$ \widehat{L}_{L}\lvert_{y=y_1}=0$, $\widehat{L}_{R}\lvert_{y=0}=0$ result in 
$
\widehat{n}_{L,R}=0
$.
Thus, imposing BC such as (\ref{VLF2}) necessarily guarantees $M_{{\mathcal L}}\neq 0$ for non-trivial solutions ($\widehat{L}_{L,R}(y)\neq0$). 
In the same way imposing the BC (\ref{VLF3}) we have $M_{\mathcal{R}}\neq 0$.  In this work we  conveniently choose $M_{L,R}(y)= -c_{L,R} W(\phi)/6$ that  results in 
a continuous spectrum for the 4D zero modes as we can see in Fig.~\ref{fig:ML}.

\section{Gauge interactions}
\label{sec:gauge}

In this section we will describe the gauge interactions of the charged [$\mathcal L(x),\ \mathcal R(x)$] and neutral [$\mathcal N(x)$] components of the VLL with zero and non-zero KK modes of gauge bosons.

\subsection*{Neutral currents}
\begin{figure}[htb]
\centering
\includegraphics[width=9cm]{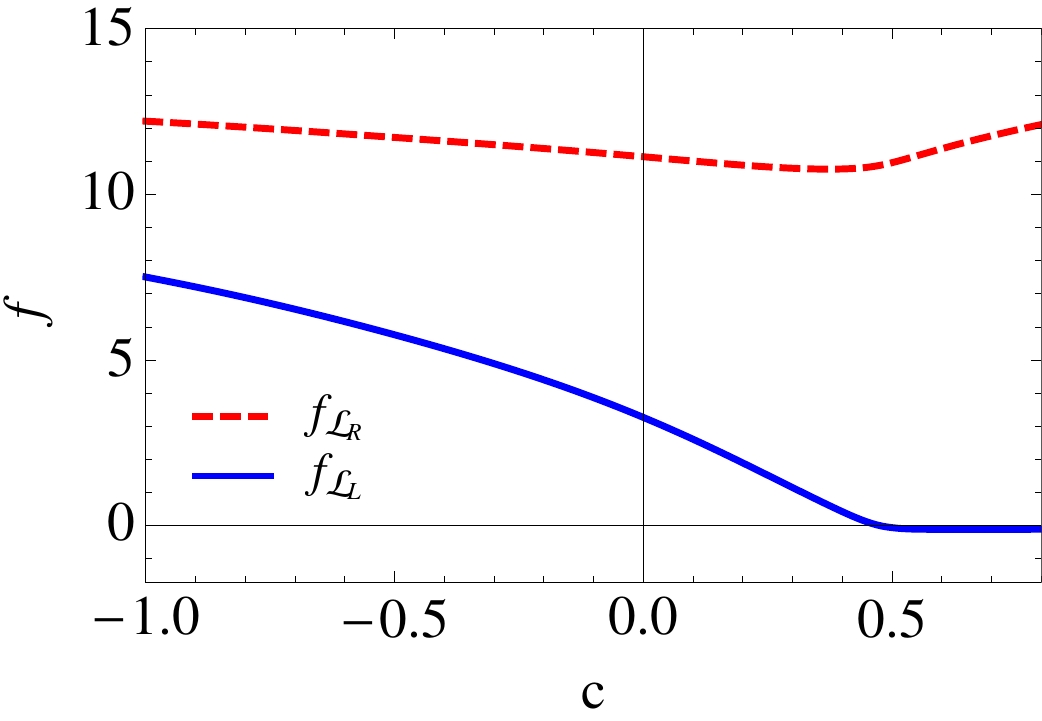} \hfill
\caption{\it Coupling of $Z_1,\gamma_1$ with VLL as a function of $c= c_{L, R}$. }
\label{fig:fLR}
\end{figure}

Before EWSB the Lagrangian describing the interactions of the charged leptons $\mu_{L}(x),\mu_R(x)$ and the charged VLL zero modes in the doublet $\mathcal L_{L,R}(x)$ and the singlet $R_{L,R}(x,y)$
$(\mu(x),\mathcal L(x),\mathcal R(x))$
with the $Z$ gauge boson and the KK modes of the gauge bosons ($Z_n,\gamma_n$), with $n\geq 1$, is given by
\begin{align}
\mathcal L=\sum_{X=Z,Z_n,\gamma_n}\mathcal L_X,\quad\mathcal L_{X}&=X_\mu\begin{pmatrix} \bar \mu_L(x) & \bar{\mathcal L}_L(x)&\bar{\mathcal R}_L(x)\end{pmatrix}\gamma^\mu G_L^X \begin{pmatrix} \mu_L(x)\\ \mathcal L_L(x)\\ \mathcal R_L(x)\end{pmatrix}\nonumber\\
&+X_\mu\begin{pmatrix} \bar \mu_R(x) & \bar{\mathcal L}_R(x)&\bar{\mathcal R}_R(x)\end{pmatrix}\gamma^\mu G_R^X \begin{pmatrix} \mu_R(x)\\ \mathcal L_R(x)\\ \mathcal R_R(x)\end{pmatrix}
\label{lagrangian}
\end{align}
where the coupling matrices $G^{X}_{L,R}$ are diagonal but not proportional to the identity
\be
G_{L,R}^X=\begin{pmatrix} g^X_{\mu_{L,R}}& 0& 0\\0&g^X_{\mathcal L_{L,R}}&0\\0&0&g^X_{\mathcal R_{L,R}}
\end{pmatrix}
\label{G}
\ee
%
%
and we will restrict ourselves to the lightest mode $n=1$, although the generalization to higher KK modes is trivial. %
The couplings in (\ref{G}) are given by
\begin{align}
g^Z_{f_2}&=\frac{1}{c_W}\left(-\frac{1}{2}+s_W^2\right)g f^Z_{f_2},\quad f_2=\mu_L,\mathcal L_{L,R}
\nonumber\\
g^{Z_n}_{f_1}&=\frac{s_W^2}{c_W}g f^{Z_n}_{f_1},\quad f_1=\mu_R,\mathcal R_{L,R}\nonumber\\
g^{\gamma_n}_{f_{1,2}}&=-s_W g f^{\gamma_n}_{f_{1,2}}
\end{align}
where $f^{X}_{f_{L,R}}$ is defined for $f=\mu,\mathcal L,\mathcal R$ as
\be
f^X_{f_{L,R}}=
\frac{{\displaystyle \sqrt{y_1} \int e^{-3A} f_X(y) f_{L,R}^2(y)}}{\left[\int f_X^2(y)\right]^{1/2}\int e^{-3A}  f_{L,R}^2(y)}\quad  (X=Z_n,\gamma_n),\quad  f^Z_{f_{L,R}}=1 \label{eq:fLR}
\ee
with $f_L(y)=\ell_L(y),L_L(y),R_L(y)$ and $f_R(y)=E_R(y),L_R(y),R_R(y)$. We show in Fig.~\ref{fig:fLR} the profile of $f^{Z_1,\gamma_1}_{{\cal L}_L}$ and $f^{Z_1,\gamma_1}_{{\cal L}_R}$ defined in Eq.~(\ref{eq:fLR}).

\subsection*{Charged currents}

The interaction Lagrangian of the neutral leptons $\mathcal N_{L,R}(x)$ with the charged leptons $\mathcal L_{L,R}(x)$ and the $W$-gauge boson and its KK excitations is given by
\begin{align}
\mathcal L_W=\sum_{n\geq 0}W_{n}^\mu(x)\left(  g^{W_n}_{\mathcal N_{L}} \bar{\mathcal N}_L(x)\gamma_\mu \mathcal L_L(x)+  g^{W_n}_{\mathcal N_{R}}\bar{\mathcal N}_R(x)\gamma_\mu \mathcal L_R(x)\right)+h.c. \label{eq:charge}
\end{align}
where
\be
 g^{W_0}_{\mathcal N_{L,R}}=\frac{g}{\sqrt{2}},\quad g^{W_n}_{\mathcal N_{L,R}}=\frac{g}{\sqrt{2}} f^W_{\mathcal N_{L,R}}\quad (n\geq 1)
\ee
and
\be
f^W_{\mathcal N_{L,R}}=
\frac{{\displaystyle \sqrt{y_1} \int e^{-3A} f_W(y) L_{L,R}(y) N_{L,R}(y)}}{\left[\int f_W^2(y)\right]^{1/2}\int e^{-3A}  L_{L,R}(y) N_{L,R}(y).   }\label{eq:WNL}
\ee
Notice that by neglecting the tiny effect of electroweak symmetry breaking, $f_{W_n}(y)=f_{Z_n,\gamma_n}(y)$, an approximation already used in the neutral current interaction.

\section{Yukawa interactions}
\label{sec:Yukawa}

We will now introduce the 5D Yukawa couplings as~\footnote{We are assuming that VLL in Eq.~(\ref{defVLF}) have lepton number $L_\mu=1$, so they can only mix through the Higgs with themselves and with the second generation leptons. Moreover the couplings between the VLL and the SM leptons could have been avoided by the simple introduction of a discrete symmetry, as in Ref.~\cite{Lee:2012xn}, an assumption we are not doing in this paper. Had we introduced it, as we will see, we would have failed to encompass the experimental value of the muon AMM.}
\begin{align}
e^{4A}\mathcal L_Y&= h(y) \left(\widehat Y_{\ell E}\bar \ell_L(x,y) E_R(x,y)+\widehat Y_{\ell R}\bar \ell_L(x,y)R_R(x,y)\right.\\
&+\left.\widehat Y_{L E}\bar L_L(x,y) E_R(x,y)+\widehat Y_{LR}(\bar L_L(x,y) R_R(x,y)+\bar R_L(x,y) L_R(x,y))\right)+h.c.\nonumber
\end{align}
where the $\widehat Ys$ are 5D Yukawa couplings with mass dimension $-1/2$. 

By expanding the 5D fermions in the KK components and keeping the zero modes, we get the 4D fermion mass matrix
\be
\mathcal L_m=\begin{pmatrix} \bar\mu_L(x) & \bar{\mathcal L}_L(x)  &\bar{\mathcal R}_L(x)\end{pmatrix}\cdot\mathcal M \cdot\begin{pmatrix}\mu_R(x)\\ \mathcal L_R(x)\\\mathcal R_R(x) \end{pmatrix}+h.c.
\ee
where
\be
\mathcal M=\begin{pmatrix}c_{\ell E} & 0 & c_{\ell R} \\
c_{L E} & M_\mathcal {L}&c_{L R} \\
0 &c_{R L} &M_{\mathcal R}    
\end{pmatrix}
\label{masa}
\ee
with entries given by
\be
c_{J K}=\widehat Y_{JK} v \frac{{\displaystyle \int e^{\alpha ky-4A} J_L(y) K_R(y) dy }}{{\displaystyle \left[ \int e^{2\alpha ky-2A} \int e^{-3A}J_L^2(y) \int e^{-3A}K_R^2(y)\right]^{1/2} }} \,, \label{eq:cJY}
\ee
for $J=\ell,L,R$ and $K=E,L,R$ with $\widehat Y_{RL}\equiv\widehat Y_{LR}$, and $v=174$ GeV.

We can now go to the mass eigenstate basis $(\mu,\mathcal L,\mathcal R)\to(\widetilde\mu,\widetilde {\mathcal L},\widetilde {\mathcal R})$ defined as
\be
\begin{pmatrix} \mu_{L,R}(x)\\ {\mathcal L}_{L,R}(x)\\   {\mathcal R}_{L,R}(x)
\end{pmatrix}=U_{L,R}\begin{pmatrix} \widetilde \mu_{L,R}(x) \\ \widetilde {\mathcal L}_{L,R}(x)\\  \widetilde {\mathcal R}_{L,R}(x)
\end{pmatrix},
\label{mixing}
\ee
where $U_{L}$ ($U_{R}$) is the unitary transformation that diagonalizes $\mathcal M \mathcal M^\dagger$ ($\mathcal M^\dagger \mathcal M$), such that the diagonalized mass matrix reads as
\be
U^\dagger_L \mathcal M U_R\equiv\diag(m_\mu, M_{\widetilde{\mathcal L}},M_{\widetilde{\mathcal R}})\ .
\ee
 %
%
%

In the same way the interaction of the fermions with the 4D Higgs field $H(x)$ can be written as~\footnote{In the limit $m_h\ll m_{KK}$ we used $\xi(x,y)=h(y)H(x)/v$.} 

\be
\mathcal L_{Hff} = H(x)\begin{pmatrix} \bar{\widetilde \mu}_L(x) & \bar{\widetilde{\mathcal L}}_L(x)&\bar{\widetilde{\mathcal R}}_L(x)\end{pmatrix} \frac{Y}{\sqrt{2}} \begin{pmatrix} \widetilde\mu_R(x)\\ \widetilde{\mathcal L}_R(x)\\ \widetilde{\mathcal R}_R(x)\end{pmatrix} +h.c.
\ee
where  the matrix of 4D Yukawa couplings $Y$ is given by  
\be
Y=\frac{1}{v}U_L^\dagger \begin{pmatrix} c_{\ell E}& 0 & c_{\ell R}\\ c_{LE}& 0 & c_{LR}\\
0 & c_{RL} & 0\end{pmatrix} U_R.
\label{Y4D}
\ee

\subsection*{Neutral currents}
From Eq.~(\ref{lagrangian}) the interactions of the $Z$ gauge boson, and the KK bosons $Z^\mu_n,\gamma^\mu_n$, with the mass eigenstates can be written as 
\begin{align}
\mathcal L=\sum_{X=Z,Z_n,\gamma_n}\mathcal L_{X},\quad\mathcal L_{X}&=X_\mu\begin{pmatrix} \bar{\widetilde \mu}_L(x) & \bar{\widetilde{\mathcal L}}_L(x)&\bar{\widetilde{\mathcal R}}_L(x)\end{pmatrix}\gamma^\mu U_L^\dagger G_L^X U_L\begin{pmatrix} \widetilde\mu_L(x)\\ \widetilde{\mathcal L}_L(x)\\ \widetilde{\mathcal R}_L(x)\end{pmatrix}\nonumber\\
&+X_\mu\begin{pmatrix} \bar {\widetilde\mu}_R(x) & \bar{\widetilde{\mathcal L}}_R(x)&\bar{\widetilde{\mathcal R}}_R(x)\end{pmatrix}\gamma^\mu U_R^\dagger G_R^X U_R\begin{pmatrix} \widetilde\mu_R(x)\\ \widetilde{\mathcal L}_R(x)\\ \widetilde{\mathcal R}_R(x)\end{pmatrix}
\end{align}
where the matrices $U_{L,R}^\dagger G_{L,R}^XU_{L,R}$ create a mixing between the muon and the VLL. 

The interaction Lagrangian with mass eigenstates involving at least one light state, $\widetilde\mu_{L,R}$, then reads as
\begin{align}
\mathcal L_{X}&=X_\mu \left(g^{X}_{\widetilde{\mu}_L} \bar {\widetilde\mu}_L\gamma^\mu \widetilde{\mu}_L+g^{X}_{\widetilde{\mathcal L}_L} \bar {\widetilde\mu}_L\gamma^\mu \widetilde{\mathcal L}_L+
g^{X}_{\widetilde{\mathcal R}_L} \bar {\widetilde\mu}_L\gamma^\mu \widetilde{\mathcal R}_L\right.\nonumber\\
&+\left. g^{X}_{\widetilde{\mu}_R} \bar {\widetilde\mu}_R\gamma^\mu \widetilde{\mu}_R+g^{X}_{\widetilde{\mathcal L}_R} \bar {\widetilde\mu}_R\gamma^\mu \widetilde{\mathcal L}_R+
g^{X}_{\widetilde{\mathcal R}_R} \bar{\widetilde \mu}_R\gamma^\mu \widetilde{\mathcal R}_R\right)+h.c.
\label{lagrangian2}
\end{align}
where the couplings with mass eigenstates are then given by
\begin{align}
g^{X}_{\widetilde{\mu}_L}=&g^X_{\mu_L}U_L^{11}U_L^{11}+
g^X_{\mathcal L_L}U_L^{21}U_L^{21}+g^X_{\mathcal R_L}U_L^{31}U_L^{31}\nonumber\\
g^{X}_{\widetilde{\mathcal L}_L}=&g^X_{\mu_L}U_L^{11}U_L^{12}+
g^X_{\mathcal L_L}U_L^{21}U_L^{22}+g^X_{\mathcal R_L}U_L^{31}U_L^{32}\nonumber\\
g^{X}_{\widetilde{\mathcal R}_L}=&g^X_{\mu_L}U_L^{11}U_L^{13}+
g^X_{\mathcal L_L}U_L^{21}U_L^{23}+g^X_{\mathcal R_L}U_L^{31}U_L^{33}\nonumber\\
g^{X}_{\widetilde{\mu}_R}=&g^X_{\mu_R}U_R^{11}U_R^{11}+
g^X_{\mathcal L_R}U_R^{21}U_R^{21}+g^X_{\mathcal R_R}U_R^{31}U_R^{31}\nonumber\\
g^{X}_{\widetilde{\mathcal L}_R}=&g^X_{\mu_R}U_R^{11}U_R^{12}+
g^X_{\mathcal L_R}U_R^{21}U_R^{22}+g^X_{\mathcal R_R}U_R^{31}U_R^{32}\nonumber\\
g^{X}_{\widetilde{\mathcal R}_R}=&g^X_{\mu_R}U_R^{11}U_R^{13}+
g^X_{\mathcal L_R}U_R^{21}U_R^{23}+g^X_{\mathcal R_R}U_R^{31}U_R^{33}
\label{acoplos}
\end{align}
%
%
%
%
and the corresponding vector and axial couplings are $g_{V,A}=\frac{1}{2}(g_L\pm g_R)$.

\subsection*{Charged currents}

From Eq.~(\ref{eq:charge}) the interaction of the neutral lepton $\mathcal N(x)$ with the physical (mass eigenstate) muon $\widetilde\mu(x)$ is given by
\be
\mathcal L_W=\sum_{n\geq 0}W_{n}^\mu(x)\left( g^{W_n}_{\widetilde\mu_{L}} \bar{\mathcal  N}_L(x)\gamma_\mu \widetilde \mu_L(x)+ g^{W_n}_{\widetilde\mu_{R}}\bar{\mathcal N}_R(x)\gamma_\mu \widetilde\mu_R(x)\right)+h.c.
\label{charged2}
\ee
where
\be
g^{W_n}_{\widetilde\mu_{L,R}}= U_{L,R}^{21} g^{W_n}_{\mathcal N_{L,R}}
\label{cargados}
\ee
and the vector and axial couplings are given by
\be
g^W_{V,A}=\frac{1}{2}\left(U_L^{21}g^W_{\mathcal N_{L}}\pm U_R^{21}g^W_{\mathcal N_{R}}   \right),\quad
\left(g^W_{V}\right)^2-\left(g^W_{A}\right)^2=U_L^{21}U_R^{21}g^W_{\mathcal N_{L}}g^W_{\mathcal N_{R}}
\ee

\section{Analytic expressions of $U_{L,R}$}

\label{sec:analytical}
 
 If the entries $c_{\ell E}$, $c_{\ell R}$ and $c_{LE}$ in the mass matrix (\ref{masa}) are much smaller than the other entries (as we will see in Sec.~\ref{sec:Zmumu} it happens in this theory), the mass matrix $\mathcal M$ can be expanded as follows:
\be
\mathcal M=\mathcal M^0+\delta M^0\equiv\begin{pmatrix}0 & 0 & 0 \\
0 & M_\mathcal {L}&c_{L R} \\
0 &c_{R L} &M_{\mathcal R}    
\end{pmatrix}+\begin{pmatrix}c_{\ell E} & 0 & c_{\ell R} \\
c_{L E} & 0 &0\\
0 &0 &0    
\end{pmatrix}
\label{Aprox1}
\ee
which will allow us to use a perturbative approach to find the matrices $U_L$,$U_R$  that diagonalize, respectively, $\mathcal M \mathcal M^\dagger$ and 
$\mathcal M^\dagger \mathcal M$. The resulting diagonalization matrices, $U_{L,R}$, are then given, to first order in the small parameters, by
\be
U_{L,R}=\begin{pmatrix}1 & U_{L,R}^{12} & U_{L,R}^{13}\\
U_{L,R}^{21} & \cos\theta_{L,R}& \sin\theta_{L,R}\\
U_{L,R}^{31} & -\sin\theta_{L,R} & \cos\theta_{L,R}
\end{pmatrix}
\label{Aprox2}
\ee
where
\begin{align}
\begin{pmatrix} U_{L,R}^{21}\\U_{L,R}^{31}\end{pmatrix}=-\begin{pmatrix}\cos\theta_{L,R}& \sin\theta_{L,R}\\ -\sin\theta_{L,R}& \cos\theta_{L,R}\end{pmatrix}\begin{pmatrix} U_{L,R}^{12}\\U_{L,R}^{13}
\end{pmatrix}
\end{align}
with
\begin{align}
U^{12}_{L}&=\frac{(c_{L R}\cos\theta_{L}-M_{R}\sin\theta_{L}) }{M^2_{\widetilde{\mathcal L}}}c_{\ell R},
&U^{13}_{L}=\frac{(c_{L R}\sin\theta_{L}+M_{ R}\cos\theta_{L})}{M^2_{\widetilde{\mathcal R}}}c_{\ell R}\nonumber\\
U^{12}_{R}&=\frac{(M_L\cos\theta_{R}-c_{LR}\sin\theta_{R})}{M^2_{\widetilde{\mathcal L}}}c_{LE},
&U^{13}_{R}=\frac{(M_L\sin\theta_{L}+c_{LR}\cos\theta_{L})}{M^2_{\widetilde{\mathcal R}}}c_{LE}
\label{Aprox3}
\end{align}
and the angles $\theta_{L,R}$ are given by
\begin{align}
\sin 2\theta_L=&\frac{2(c_{R L}M_{\mathcal L}+c_{L R}M_{\mathcal R})}{|M^2_{\widetilde{\mathcal L}}-M^2_{\widetilde{\mathcal R}}|}\nonumber\\
\sin 2\theta_R=&\frac{2(c_{L R}M_{\mathcal L}+c_{R L}M_{\mathcal R})}{|M^2_{\widetilde{\mathcal L}}-M^2_{\widetilde{\mathcal R}}|}.
\label{Aprox6}
\end{align}
In this approximation the mass eigenvalues are then given by
\begin{align}
m_\mu&=c_{\ell E}+\frac{c_{\ell R}
c_{L E}}{M_{\widetilde{\mathcal L}}M_{\widetilde{\mathcal R}}}\,c_{R L}\nonumber\\
M^2_{\widetilde{\mathcal L}, \widetilde{\mathcal R}}&=\frac{1}{2}\Bigg \{M^2_{\mathcal L}+M^2_{\mathcal R}+c^2_{R L}+c^2_{L R}\nonumber\\
&\mp\sqrt{\left((M_{\mathcal L}+M_{\mathcal R})^2+(c_{L R}-c_{R L})^2\right)\left((M_{\mathcal L}-M_{\mathcal R})^2+(c_{L R}+c_{R L})^2\right)}\Bigg\}
\label{Aprox7}
\end{align} 

From Eq.~(\ref{acoplos}) the interactions of the $Z$ gauge boson and the neutral KK bosons $Z^\mu_n,\gamma^\mu_n$ with the mass eigenstates can be written as
 \begin{align}
g^{X}_{\widetilde{\mu}_L}=&g^X_{\mu_L}+
g^X_{\mathcal L_L}U_L^{21}U_L^{21}+g^X_{\mathcal R_L}U_L^{31}U_L^{31}\nonumber\\
g^{X}_{\widetilde{\mu}_R}=&g^X_{\mu_R}+
g^X_{\mathcal L_R}U_R^{21}U_R^{21}+g^X_{\mathcal R_R}U_R^{31}U_R^{31}\nonumber\\
g^{X}_{\widetilde{\mathcal L}_L}=&\cos\theta_{L}(g^X_{\mathcal L_L}-g^X_{\mu_L})U_L^{21}-\sin\theta_{L}(g^X_{\mathcal R_L}-g^X_{\mu_L})U_L^{31}\nonumber\\
g^{X}_{\widetilde{\mathcal L}_R}=&\cos\theta_{R}(g^X_{\mathcal L_R}-g^X_{\mu_R})U_R^{21}-\sin\theta_{R}(g^X_{\mathcal R_R}-g^X_{\mu_R})U_R^{31}\nonumber\\
g^{X}_{\widetilde{\mathcal R}_L}=&\sin\theta_{L}(g^X_{\mathcal L_L}-g^X_{\mu_L})U_L^{21}+\cos\theta_{L}(g^X_{\mathcal R_L}-g^X_{\mu_L})U_L^{31}\nonumber\\
g^{X}_{\widetilde{\mathcal R}_R}=&\sin\theta_{R}(g^X_{\mathcal L_R}-g^X_{\mu_R})U_R^{21}+\cos\theta_{R}(g^X_{\mathcal R_R}-g^X_{\mu_R})U_R^{31}.
\label{Aprox8}
\end{align}
Notice that $$g^{X}_{\widetilde{\mu}_{L,R}}=g^X_{\mu_{L,R}}+\cdots$$ where the ellipsis denotes  (subleading) terms which are quadratic in the small perturbations.

The interactions of the $W$ gauge boson and the charged KK modes $W_n$ with the mass eigenstates in terms of the elements of $U_{L,R}$ were already given in Eq.~(\ref{cargados}).

\section{The case $c_L=c_R\equiv c$}
\label{sec:cLcR}

As we have seen in the previous sections the mass eigenvalues and mixing angles in the muon/VLL sector depend on the two real parameters $c_L$ and $c_R$, which determine the localization along the extra dimension, respectively, of the doublet and singlet VLL. 
In this section we will consider the particularly simple case where  $c_L=c_R\equiv c$~\footnote{The general case $c_L\neq c_R$ can be worked out straightforwardly.}.
In this
case we will be able to write all the functions $c_{J K}$  in term of the function $c_{L R}$
and the parameter $c$. 
This will be achieved by setting  values for the elements $U^{31}_{L},U^{21}_{R}$ in order to satisfy the experimental bounds
for $\delta g^Z_{\mu_{L,R}}$, as we will see in Eq.~(\ref{bounds}).

For $c_R=c_L$ we have $M_{\mathcal L}(c)=M_{\mathcal R}(c)\equiv M(c)$ and the 5D wave functions are related to each other as
$N_{L,R}(y)=L_{L,R}(y)= R_{L,R}(y)$ resulting in  
 \begin{align}
c_{L R}=c_{R L} \textrm{ and } \theta_{L}=\theta_{R}=\pi/4.
\label{valores}
\end{align}
Equation~(\ref{valores}) and the explicit form for the matrix $U_{L,R}$  
enable us to write 
 \begin{align}
 M_{\widetilde{\mathcal  L}}(c,c_{LR})&=M(c)-c_{L R}\nonumber\\
M_{\widetilde{\mathcal  R}}(c,c_{LR})&=M(c)+c_{L R}
 \label{case1}
\end{align}
\begin{figure}[htb]
\centering
\includegraphics[width=7.55cm]{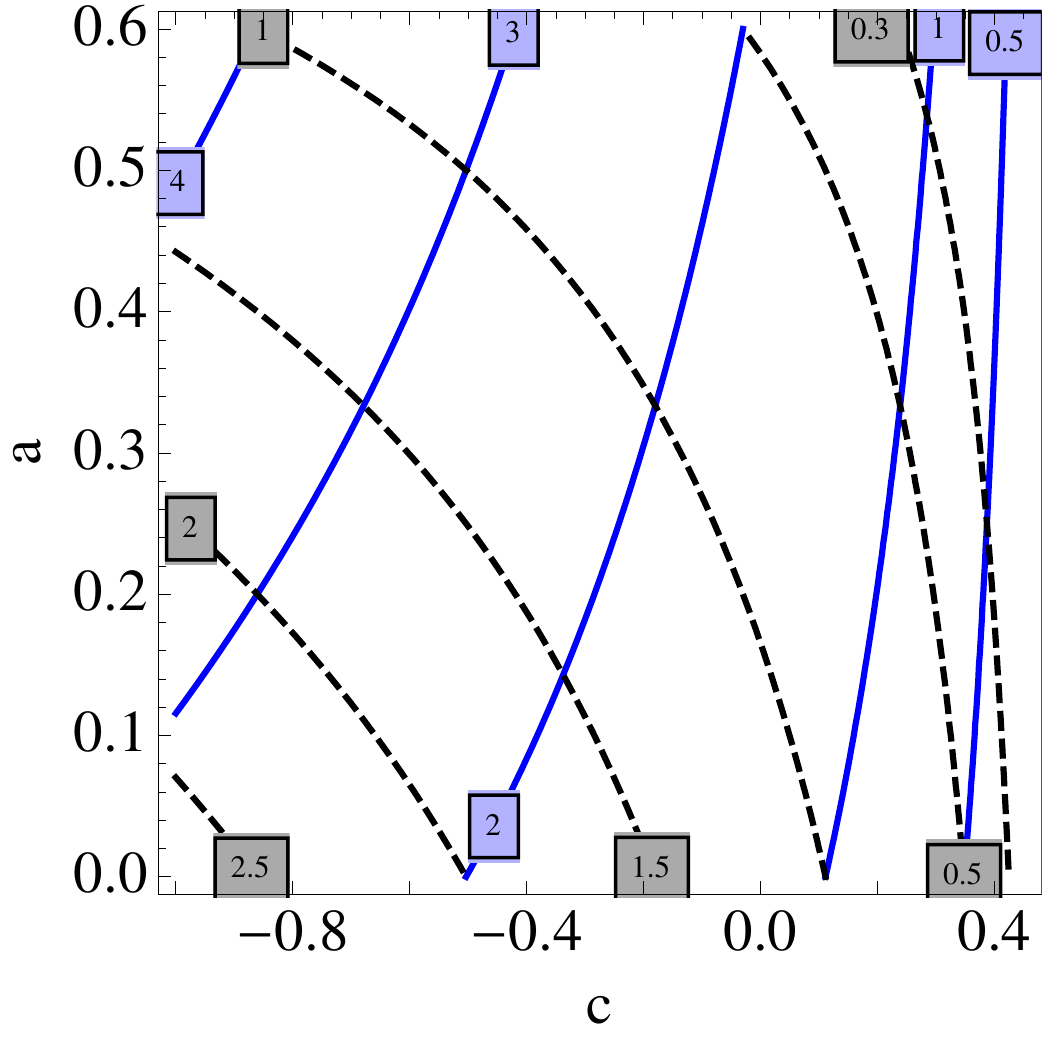} \hfill
\includegraphics[width=7.55cm]{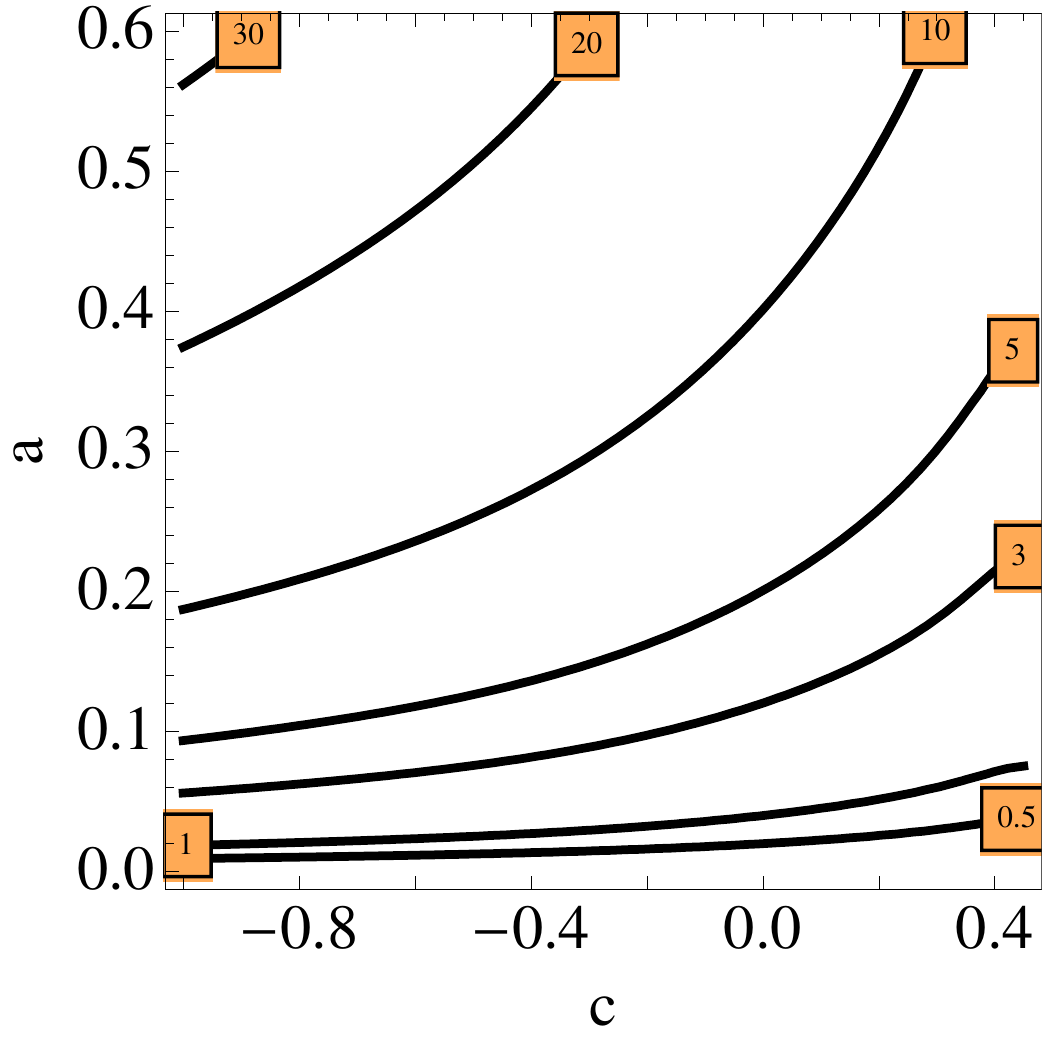}
\caption{\it Left panel: Contour plot of $M_{\widetilde{\mathcal L}}$ (dashed black) and $M_{\widetilde{\mathcal R}}$ (solid blue) in the plane $(c,a)$. The labels of the contours are in $\textrm{TeV}$. Right panel: Contour plot of Yukawa coupling in 5D, $\sqrt{k} \widehat Y_{LR}$ in the plane $(c,a)$. We have considered  $c_{\mu_L}=0.4$, $c_{\mu_R}=0.5$.}
\label{fig:MLR}
\end{figure} 
and
 \begin{align}
 c_{\ell E} &=m_\mu- 2\frac{M_{\widetilde{\mathcal  R}}M_{\widetilde{\mathcal  L}}(M_{\widetilde{\mathcal R}}-M_{\widetilde{\mathcal L}})}{(M_{\widetilde{\mathcal  R}}+M_{\widetilde{\mathcal  L}})^2}U_L^{31}U_R^{21}\nonumber\\
c_{\ell R}&=-2 \frac{M_{\widetilde{\mathcal  R}}M_{\widetilde{\mathcal  L}}}{M_{\widetilde{\mathcal  R}}+M_{\widetilde{\mathcal  L}}} U_L^{31}\nonumber\\
c_{L E}&=-2 \frac{M_{\widetilde{\mathcal  R}}M_{\widetilde{\mathcal  L}}}{M_{\widetilde{\mathcal  R}}+M_{\widetilde{\mathcal  L}}} U_R^{21}.
 \label{case2}
\end{align}
Then setting  values for the elements $U^{31}_{L},U^{21}_{R}$ consistently with experimental bounds we can write all the functions $c_{J K}$  in term of the parameters $c$ and the function $c_{L R}$. Moreover by defining the parameter $\beta$ as
\be 
\beta=-\frac{M_{\widetilde{\mathcal  R}}-M_{\widetilde{\mathcal  L}}}{M_{\widetilde{\mathcal  R}}+M_{\widetilde{\mathcal  L}}}\label{case3}
\ee
we get
\be 
U_L^{21}=\beta U_L^{31},\quad  U_R^{31}=\beta U_R^{21}.
\ee

We will change the independent parameters from $(c,c_{LR})$ to $(c,a)$ by introducing the convenient parametrization 
\be
a\equiv c_{LR}/M(c)
\label{adef}
\ee
and present the results for the mass eigenvalues in the plane $(c,a)$ in the left panel of Fig.~\ref{fig:MLR}. In the absence of a theory predicting the 5D Yukawa couplings we will consider them as output from the different constraints. In particular, using the variable $a$ from Eq.~(\ref{adef}), implies an implicit assumption for the 5D Yukawa coupling $\widehat Y_{LR}$. The required values of $\widehat Y_{LR}$ are shown as contour plots in the plane $(c,a)$ in the right panel of Fig.~\ref{fig:MLR}. The condition of perturbativity of the 5D theory would imply an upper bound on the 5D Yukawa couplings such that $\sqrt{k} \widehat{Y}_{LR}\lesssim 4\pi$ which already excludes the upper left corner of the parameter region in the plane $(c,a)$, as we can see from the right panel of Fig.~\ref{fig:MLR}. Nevertheless this region, as we will see, is also excluded by electroweak constraints which in fact rule out the region $\sqrt{k} \widehat{Y}_{LR}\gtrsim 4$.

Gauge and Yukawa couplings, in the particular case we are considering in this section, also take simplified values which we now describe.

\subsection{Gauge couplings}

The couplings with the $Z$ gauge boson and the neutral KK bosons $Z_n,\gamma_n$ can be written, from Eq.~(\ref{Aprox8}), as
 \begin{align}
g^{X}_{\widetilde{\mathcal L}_L}=&\frac{U_L^{31}}{\sqrt{2}}\left[\beta(g^X_{\mathcal L_L}-g^X_{\mu_L})-(\tan\alpha_X g^X_{\mathcal L_L}-g^X_{\mu_L} )\right]\nonumber\\
g^{X}_{\widetilde{\mathcal L}_R}=&\frac{U_R^{21}}{\sqrt{2}}\left[(g^X_{\mathcal L_R}-g^X_{\mu_R})-\beta(\tan\alpha_X g^X_{\mathcal L_R}-g^X_{\mu_R} )\right]\nonumber\\
g^{X}_{\widetilde{\mathcal R}_L}=&\frac{U_L^{31}}{\sqrt{2}}\left[\beta(g^X_{\mathcal L_L}-g^X_{\mu_L})+(\tan\alpha_X g^X_{\mathcal L_L}-g^X_{\mu_L} )\right]\nonumber\\
g^{X}_{\widetilde{\mathcal R}_R}=&\frac{U_R^{21}}{\sqrt{2}}\left[(g^X_{\mathcal L_R}-g^X_{\mu_R})+\beta(\tan\alpha_X g^X_{\mathcal L_R}-g^X_{\mu_R} )\right]
\label{case4}
\end{align}
with
\be 
\tan\alpha_X =\left\{\begin{array}{lcl}
 \frac{g^{Z,SM}_{\mu_R}}{g^{Z,SM}_{\mu_L}}  & \mbox{if} &  X=Z,\,Z_n\\
1 &\mbox{if}& X=\gamma_n.
\label{case33}
\end{array}\right. 
\ee
where $g_{\mu_{L,R}}^{Z,SM}$ denotes the SM (tree-level) $Z$ coupling to the $\mu_{L,R}$ fields. 

On the other hand, the couplings with the $W$ gauge boson and the charged KK bosons $W_n$, Eq.~(\ref{cargados}), are written as
\be
g^{W_n}_{\widetilde\mu_{L}}= \beta U_{L}^{31} g^{W_n}_{N_{L}},\quad
g^{W_n}_{\widetilde\mu_{R}}= U_{R}^{21} g^{W_n}_{N_{R}},\quad \left(g^{W_n}_{V}\right)^2-\left(g^{W_n}_{A}\right)^2=\beta U_L^{31}U_R^{21}g^{W_n}_{N_{L}}g^{W_n}_{N_{R}}\quad (n\geq 0)
\label{cargadosfinal}
\ee
\subsection{Yukawa couplings}

We can also write the 4D Yukawa couplings of the Higgs with mass eigenstates in Eq.~(\ref{Y4D}) as
\be
Y=\frac{1}{v} \begin{pmatrix} c_{\ell E}& \frac{M_{\widetilde{\mathcal  R}}U^{31}_L}{
\sqrt{2}} & -\frac{M_{\widetilde{\mathcal  L}}U^{31}_L}{\sqrt{2}}\\ 
-\frac{M_{\widetilde{\mathcal  R}}U^{21}_R}{\sqrt{2}}& -(\frac{M_{\widetilde{\mathcal  R}}-M_{\widetilde{\mathcal L}}}{2}) & 0\\
-\frac{M_{\widetilde{\mathcal  L}}U^{21}_R}{\sqrt{2}} & 0 & \frac{M_{\widetilde{\mathcal  R}}-M_{\widetilde{\mathcal  L}}}{2}\end{pmatrix} +\cdots
\label{Y4D2}
\ee
where the ellipsis refers to terms which are subleading (quadratic) in the small parameters $U_L^{31}$ and $U_R^{21}$. As we will see below the tiny elements $Y^{12},Y^{21}$ and $Y^{13},Y^{31}$ will generate small (subleading) corrections to the muon anomalous magnetic moments while the diagonal elements $Y^{22}$ and $Y^{33}$ will contribute to the Higgs branching fractions of $H\to\gamma\gamma$ and will constrain the parameter space, or can be an indirect measurement of VLL if in the future there is an excess of $\gamma\gamma$ events. 

\section{$\delta g_{Z\overline{\mu}\mu}$ from VLL} 
\label{sec:Zmumu}
As VLL mix with the muon sector, the most important effect of the presence of VLL is the modification of the coupling of the $Z$ gauge boson with the physical muon. We have singled out this constraint as it will unambiguously determine part of the theory parameters. In particular we will see that it determines the size of the relevant mixing parameters $U_L^{31}$ and $U_R^{21}$.

In the presence of the mixing (\ref{mixing}) the SM coupling of the $Z$ gauge boson with muons $g^{Z,SM}_{\mu_{L,R}}$ gets modified. In fact we have defined the coupling matrix to $Z$ gauge bosons in Eq.~(\ref{acoplos}). We can assume here, to leading order, that $f_Z(y)=1$ so that the coupling matrices can be written as
\begin{align}
G_L^Z&=G_L^{Z,SM}+\delta G_L^Z = g^{Z,SM}_{\mu_L}\left[\mathbb I_3+\diag\left(0,0,\frac{1}{2 s^2_W-1}\right)\right]\nonumber\\
G_R^Z&=G_R^{Z,SM}+\delta G_R^Z = g^{Z,SM}_{\mu_R}\left[\mathbb I_3+\diag\left(0,-\frac{1}{2 s^2_W},0\right)\right]
\end{align}
Going now to the mass eigenstates as in Eq.~(\ref{lagrangian}) 
\begin{align}
\mathcal L_{Z}&=Z_\mu\begin{pmatrix} \overline{\widetilde \mu}_L(x) & \overline{\widetilde{\mathcal L}}_L(x)&\overline{\widetilde{\mathcal R}}_L(x)\end{pmatrix}\gamma^\mu U_L^\dagger G_L^Z U_L\begin{pmatrix} \widetilde\mu_L(x)\\ \widetilde{\mathcal L}_L(x)\\ \widetilde{\mathcal R}_L(x)\end{pmatrix}\nonumber\\
&+Z_\mu\begin{pmatrix} \overline {\widetilde\mu}_R(x) & \overline{\widetilde{\mathcal L}}_R(x)&\overline{\widetilde{\mathcal R}}_R(x)\end{pmatrix}\gamma^\mu U_R^\dagger G_R^Z U_R\begin{pmatrix} \widetilde\mu_R(x)\\ \widetilde{\mathcal L}_R(x)\\ \widetilde{\mathcal R}_R(x)\end{pmatrix}
\end{align}
we can write
\begin{align}
\frac{\delta g^Z_{\mu_L}}{g^Z_{\mu_L}}&=-\frac{(U_L^{31})^2}{1-2s^2_W},\nonumber\\
\frac{\delta g^Z_{\mu_R}}{g^Z_{\mu_R}}&=-\frac{(U_R^{21})^2}{2s^2_W}.\label{bounds}
\end{align}
Using now the experimental bound $\left|\delta g^Z_{\mu_{L,R}}/g^Z_{\mu_{L,R}}\right|\lesssim  10^{-3}$~\cite{Agashe:2016kda} we obtain for the relevant entries the upper bounds 
\be
|U_L^{31}|,\, |U_R^{21}|\lesssim 0.02
\label{Ubounds}
\ee

\begin{figure}[htb]
\centering
\includegraphics[width=5.2cm]{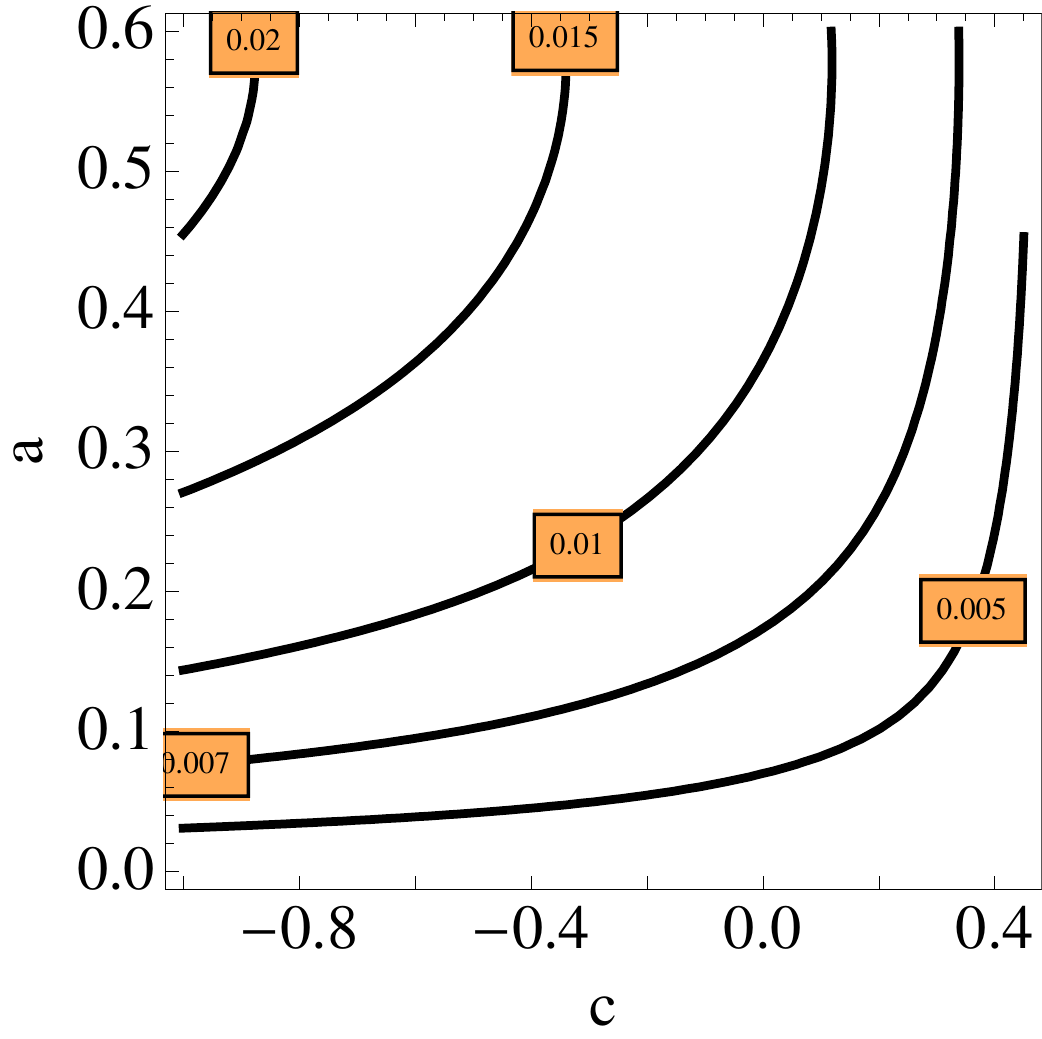} 
\includegraphics[width=5.2cm]{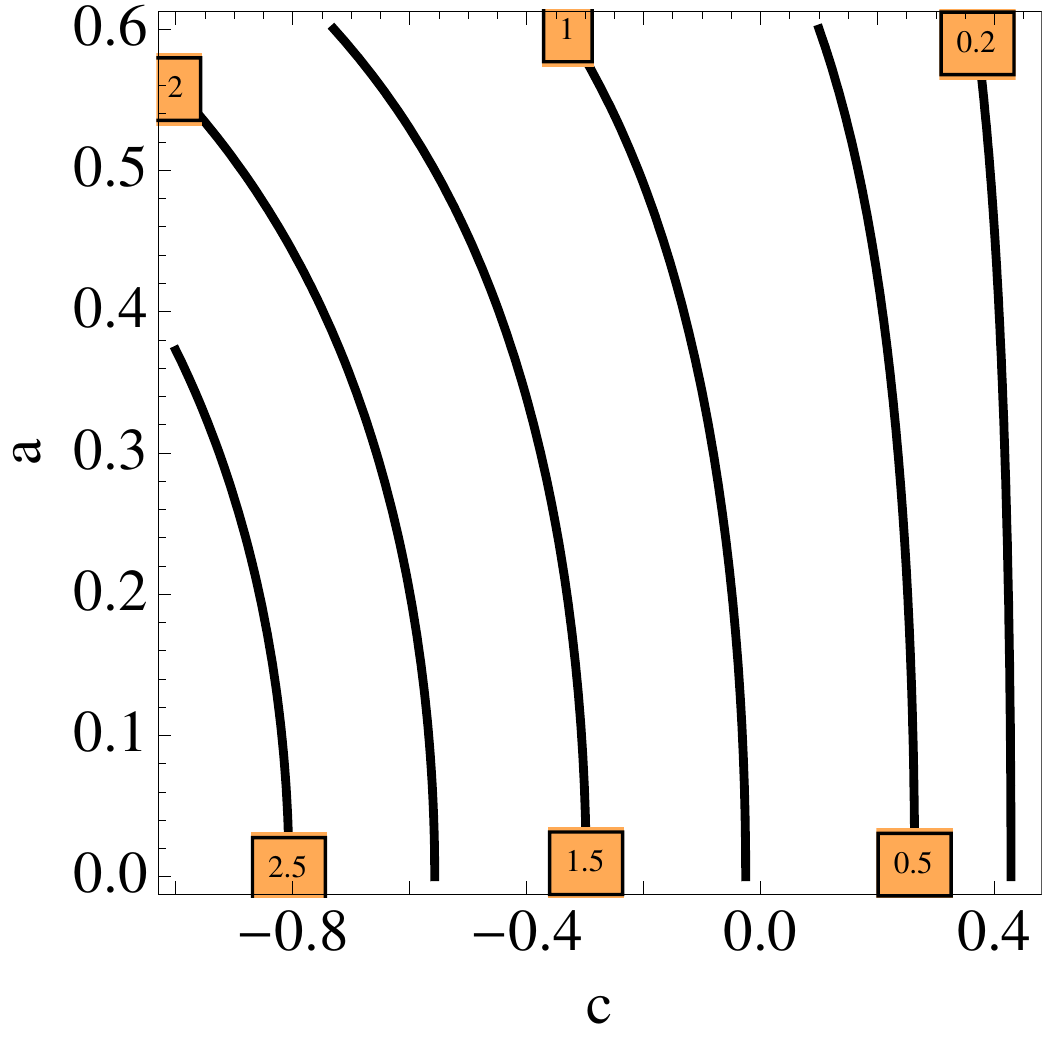} 
\includegraphics[width=5.2cm]{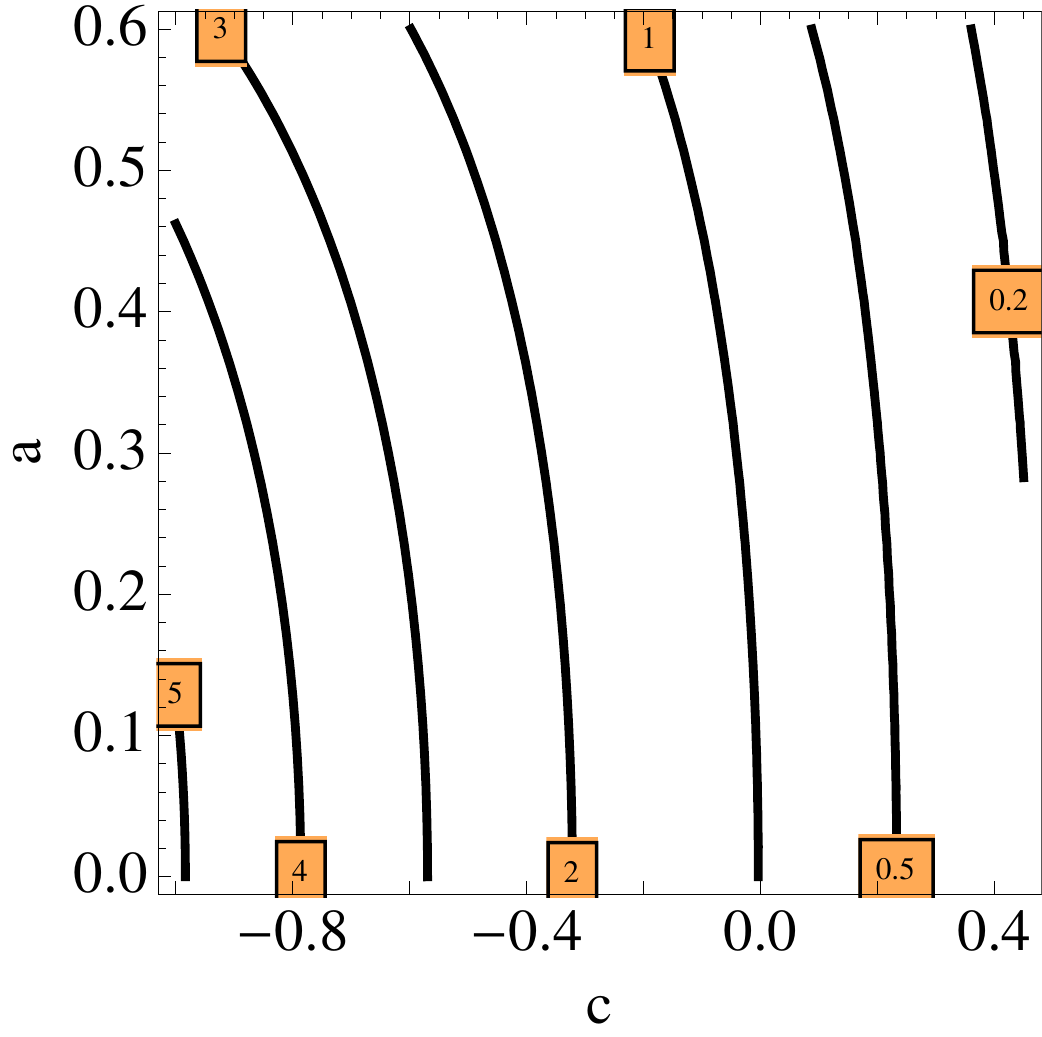} 
\caption{\it Contour plot of the 5D Yukawa couplings,  $\sqrt{k} \widehat Y_{\ell E}$ (left panel)   $\sqrt{k} |\widehat Y_{\ell R}|$ (middle panel), and $\sqrt{k}\widehat Y_{L E}$ (right panel) in the plane $(c,a)$. We have considered  $c_{\mu_L}=0.4$, $c_{\mu_R}=0.5$.}
\label{fig:Y5D1}
\end{figure} 
Using these values we present in Fig.~\ref{fig:Y5D1} the values for the Yukawa couplings $\widehat Y_{\ell E}$ (left panel), $\widehat Y_{\ell R}$ (middle panel) and $\widehat Y_{LR}$ (right panel) from Eq.~(\ref{case2}). A first observation is that the three 5D Yukawa couplings are in the perturbative regime for all values of the parameters in the $(c,a)$ plane. In fact  $\sqrt{k}\widehat Y_{\ell R},\sqrt{k}\widehat Y_{LR}\lesssim 4$ while $0.005\lesssim \sqrt{k}\widehat Y_{\ell E}\lesssim 0.02$. Moreover the small values of $\sqrt{k}Y_{\ell E}$ imply a certain degree of fine-tuning, as the mechanism to give mass to the muon in this model is somewhat different from the usual mechanism to give masses to fermions in Randall-Sundrum-like models (by means of different localizations in the extra dimension and anarchic $\mathcal O(1)$ 5D Yukawa couplings). In our case the muon mass is fixed by the first line in Eq.~(\ref{case2}), where the second term on the right-hand side is typically of $\mathcal O$(GeV), as it has to be the left-hand side term, whose  small Yukawa coupling pre-factor comes from the degree of compositeness of the muon as required in order to fit the LHCb anomaly. Finally the typical fine-tuning between both $\mathcal O$(GeV) terms, to yield the physical muon mass ($\sim 0.1$ GeV) is then expected to be $\sim 10\%$.

\section{$\Delta a_\mu$ from VLL}

\label{sec:Deltaamu}

Charged ($\widetilde{\mathcal L},\, \widetilde{\mathcal R}$) and neutral ($\mathcal N$) vector like fermions contribute to the muon anomalous magnetic moment. Charged VLL make use of neutral current 
interactions with $Z,\,Z_n,\,\gamma_n,\,H$, and neutral ones make use of charged current interactions with $W,\,W_n$. They will therefore provide corresponding contributions to the
muon AMM.

\subsection{Charged VLL}

\begin{figure}[htb!]
\begin{center}
\includegraphics[width=10cm]{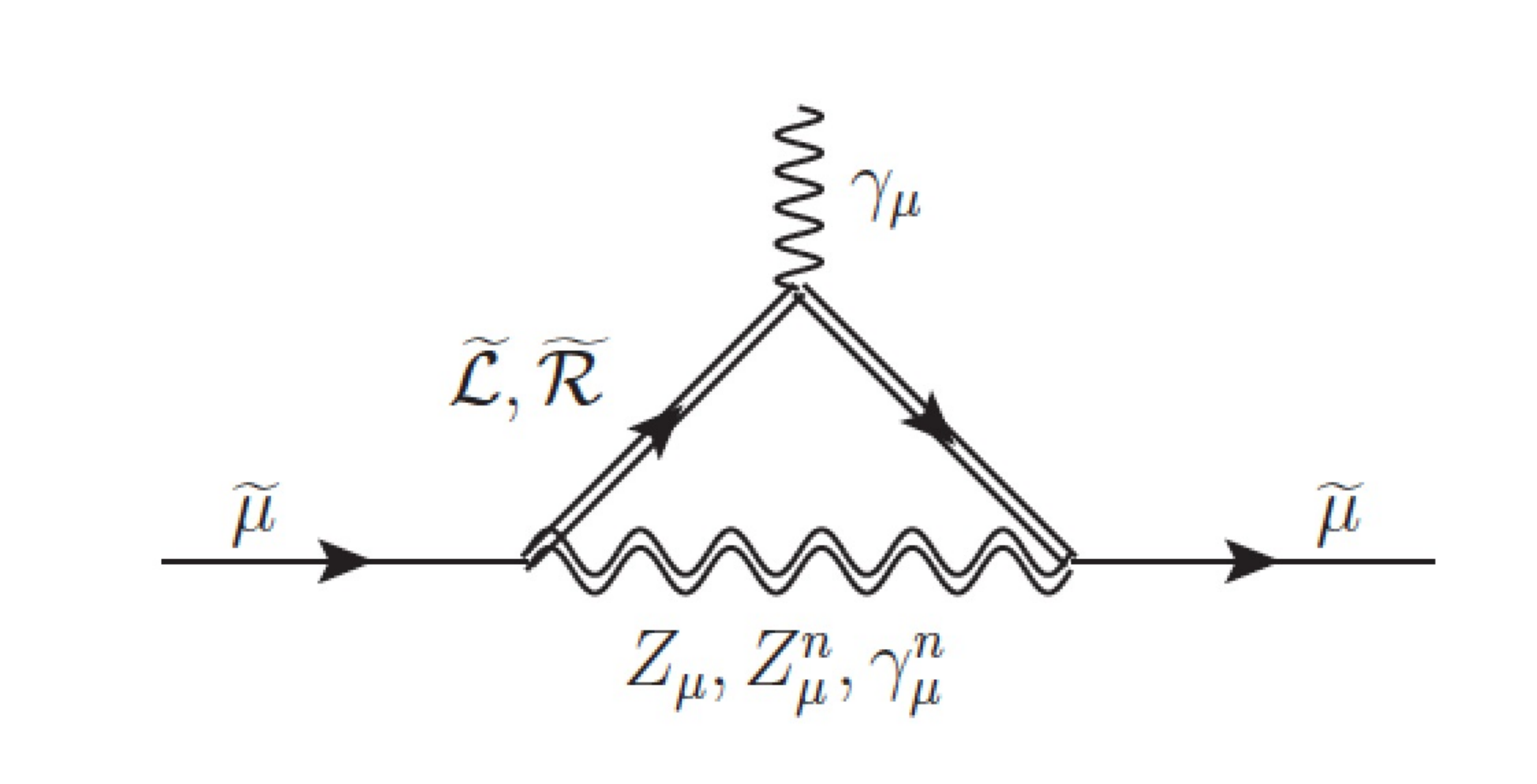} \hfill
\caption{\it Diagrams contributing to $\Delta a^C_\mu$ from charged VLL. }
\label{fig:Delta-amu-Z}
\end{center}
\end{figure}

Charged vector like fermions contribute in loops to the muon AMM as shown in Fig.~\ref{fig:Delta-amu-Z}.
%
%
%
The relevant quantity for their contribution to $\Delta a_\mu
$ is, in a very good approximation~\footnote{We are not considering subleading contributions coming from KK modes of VLL and from higher ($n\geq 2$) KK modes.}, given by 
\cite{Queiroz:2014zfa,Freitas:2014pua}
\be
\Delta a_\mu^C=\sum_{X=Z,Z^n,\gamma^n} \Delta a_\mu^X,\quad  \Delta a_\mu^X\equiv\frac{1}{4\pi^2}K_X\quad 
\ee
where
\be
K_X=\sum_{f=\widetilde{\mathcal L},\widetilde {\mathcal R}}\frac{m_\mu^2}{m_X^2}\left[ \left( g^X_{f_V} \right)^2-\left( g^X_{f_A} \right)^2 \right]\frac{M_f}{m_\mu}F_0\left(\frac{M_f}{m_X}\right)
\ee
and the function $F_0(x)$, given by
\be
F_0(x)=\frac{1-(3/4) x^2-(1/4)x^6+3 x^2 \log x}{(1-x^2)^3},
\ee
is a monotonously decreasing function such that $F_0(0)=1$ and $F_0(\infty)=1/4$. 
Using the couplings in (\ref{acoplos}) we can write
\be
K_X=\frac{m_\mu^2}{m_X^2}\left[g^X_{\widetilde{\mathcal L}_L} g^X_{\widetilde{\mathcal L}_R} \frac{M_{\widetilde{\mathcal  L}}}{m_\mu}F_0\left( \frac{M_{\widetilde{\mathcal L}}}{m_X}\right)+ g^X_{\widetilde{\mathcal R}_L} g^X_{\widetilde{\mathcal R}_R}\frac{M_{\widetilde{\mathcal R}}}{m_\mu} F_0\left( \frac{M_{\widetilde{\mathcal R}}}{m_X}\right)\right]\,.
\ee

In order to find a more explicit expression for $\Delta a_\mu$, as function of $(c,a)$, 
we first consider the contribution of KK-modes $X=Z_n,\gamma_n$ when  $g^X_{\mathcal L_L}>>g^X_{\mu_L}$ and $g^X_{\mathcal L_R} >> g^X_{\mu_R}$. In this case we can write 
 \begin{align}
g^{X}_{\widetilde{\mathcal L}_L}g^{X}_{\widetilde{\mathcal L}_R}=&\frac{U_R^{21}U_L^{31}g^X_{\mathcal L_L}g^X_{\mathcal L_R}
(\beta-\tan\alpha_X)(1-\beta\tan\alpha_X)}{2}\nonumber\\
g^{X}_{\widetilde{\mathcal R}_L}g^{X}_{\widetilde{\mathcal R}_R}=&\frac{U_R^{21}U_L^{31}g^X_{\mathcal L_L}g^X_{\mathcal L_R}
(\beta+\tan\alpha_X)(1+\beta\tan\alpha_X)}{2}
\label{case5}
\end{align}
and
 \begin{align}
\Delta a^X_\mu&=\frac{m_\mu U_R^{21}U_L^{31}g^X_{\mathcal L_L}g^X_{\mathcal L_R}}{8\pi^2 m_X^2} \nonumber\\
 &\times \Bigg\{ \beta (1+\tan^2\alpha_X)\left[M_{\widetilde{\mathcal L}} F_0\left( \frac{M_{\widetilde {\mathcal L}}}{m_X}\right)+M_{\widetilde{\mathcal R}} F_0\left( \frac{M_{\widetilde {\mathcal R}}}{m_X}\right)\right]\nonumber\\
&-  \tan\alpha_X (1+\beta^2)\left[M_{\widetilde{\mathcal L}} F_0\left( \frac{M_{\widetilde{\mathcal L}}}{m_X}\right)-M_{\widetilde{\mathcal R}} F_0\left( \frac{M_{\widetilde{\mathcal R}}}{m_X}\right)\right] \Bigg\}\,.
\label{case55}
\end{align}
On the other hand the contribution from the $Z$ gauge boson is given by
\be
\Delta a^Z_\mu=\frac{m_\mu}{4\pi^2 m_Z^2}\frac{U_L^{31}U_R^{21}}{2}(g^Z_{\mu_L}-g^Z_{\mu_R})^2\left[M_{\widetilde{\mathcal L}} F_0\left( \frac{M_{\widetilde{\mathcal L}}}{m_Z}\right)-M_{\widetilde{\mathcal R}} F_0\left( \frac{M_{\widetilde{\mathcal R}}}{m_Z}\right)\right]
\ee
Notice that the contribution from the $Z$ gauge boson is important as the relative enhancement ($\propto m_X^2/m_Z^2$) in $\Delta a_\mu$ is not compensated by the small Standard Model couplings.

Finally the Yukawa interactions in Eq.~(\ref{Y4D2}) generate an extra contribution to $\Delta a_\mu$ mediated by the diagram of Fig.~\ref{fig:Delta-amu-Z}, where the line propagating gauge and 
KK bosons is replaced~\footnote{Here we again neglect the tiny contribution from ($n\geq 1$) KK modes of the Higgs boson.}  by the Higgs propagator. The result is provided by the general expression~\cite{Queiroz:2014zfa,Freitas:2014pua}
\be
\Delta a_\mu^H=\sum_{f=\widetilde{\mathcal L},\widetilde {\mathcal R}}\frac{m_\mu^2}{192\pi^2 M_f^2}\left[ \left( Y_{\widetilde\mu_L f_R} \right)^2 + \left( Y_{\widetilde\mu_R f_L} \right)^2 \right]F_2\left(\frac{M^2_f}{m^2_H}\right)
\ee
where
\be
F_2(x)=\frac{x^4-6 x^3+3x^2 +2x+ 6x^2\log x}{(x-1)^4}\,.
\ee

For the case $c_L=c_R\equiv c$, using Eq.~(\ref{Y4D2}) yields the result
\be
\Delta a_\mu^H=\frac{m_\mu^2}{384\pi^2 v^2}\left[ \left( U^{31}_L \right)^2+\left( U^{21}_R \right)^2 \right]
\left[ \frac{M^2_{\widetilde{\mathcal  R}}}{M^2_{\widetilde{\mathcal  L}}}F_2\left(\frac{M^2_{\widetilde{\mathcal  L}}}{m^2_H}\right)+ 
\frac{M^2_{\widetilde{\mathcal L}}}{M^2_{\widetilde{\mathcal  R}}}F_2\left(\frac{M^2_{\widetilde{\mathcal  R}}}{m^2_H}\right)\right]\,.
\ee

Notice that due to the strong suppression on the values of $U_L^{31}$ and $U_R^{21}$ in Eq.~(\ref{Ubounds}) the Higgs contribution will be subleading, thus not contributing significantly to the $g_\mu-2$ anomaly.

\subsection{Neutral VLL}

\begin{center}
\begin{figure}[htb!]
\vspace{-1cm}
\hspace{4cm}\includegraphics[width=9cm]{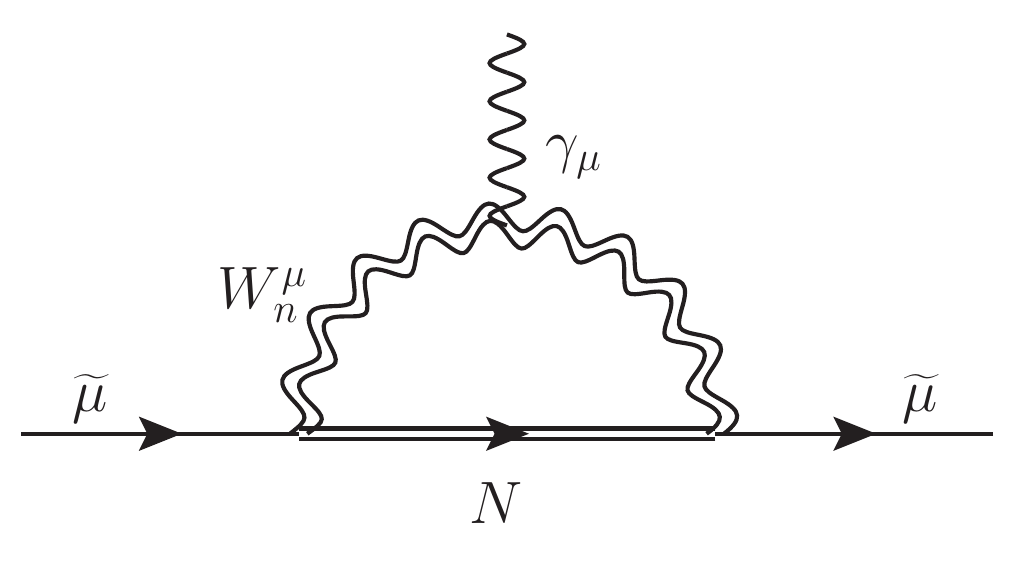} \hfill
\caption{\it Diagrams contributing to $\Delta a^N_\mu$ from neutral VLL. }
\label{fig:Delta-amu-W}
\end{figure} 
\end{center}

Moreover the contribution of the neutral vector like fermion $\mathcal N$ in loops to the anomalous magnetic moment of the muon $\Delta a_\mu^\mathcal N$ comes from the diagrams in Fig.~\ref{fig:Delta-amu-W}.
Similarly to the previous section we can now write $\Delta a_\mu^\mathcal N$ as
\be
\Delta a_\mu^\mathcal N=\frac{1}{4\pi^2}\sum_{n\geq 0}K_{W_n}
\ee
with
\be
K_{W_n}=\frac{m_\mu^2}{m_{W_n}^2} g_{\widetilde\mu_L}^{W_n}g_{\widetilde\mu_R}^{W_n}\frac{M}{m_\mu}F_1\left(\frac{M}{m_{W_n}} \right)
\ee
where  $M=M(c)$ is the mass of the neutral VLL ($\mathcal N$) and
\be
F_1(x)=\frac{-1+(17/4) x^2-3 x^4-(1/4)x^6+5 x^4\log x}{(1-x^2)^3}\,.
\ee

Using now the results in Eq.~(\ref{cargados}) we can write
\be
\Delta a_\mu^\mathcal N=\sum_{n\geq 0}\frac{m_\mu U_R^{21} U_L^{31} g^{W_n}_{N_L}g^{W_n}_{N_R}}{4\pi^2 m_{W_n}^2}
\beta  M F_1\left(\frac{M}{m_{W_n}}  \right)\,.
\ee
\subsection{Numerical results}
\begin{figure}[htb]
\centering
\includegraphics[width=8.5cm]{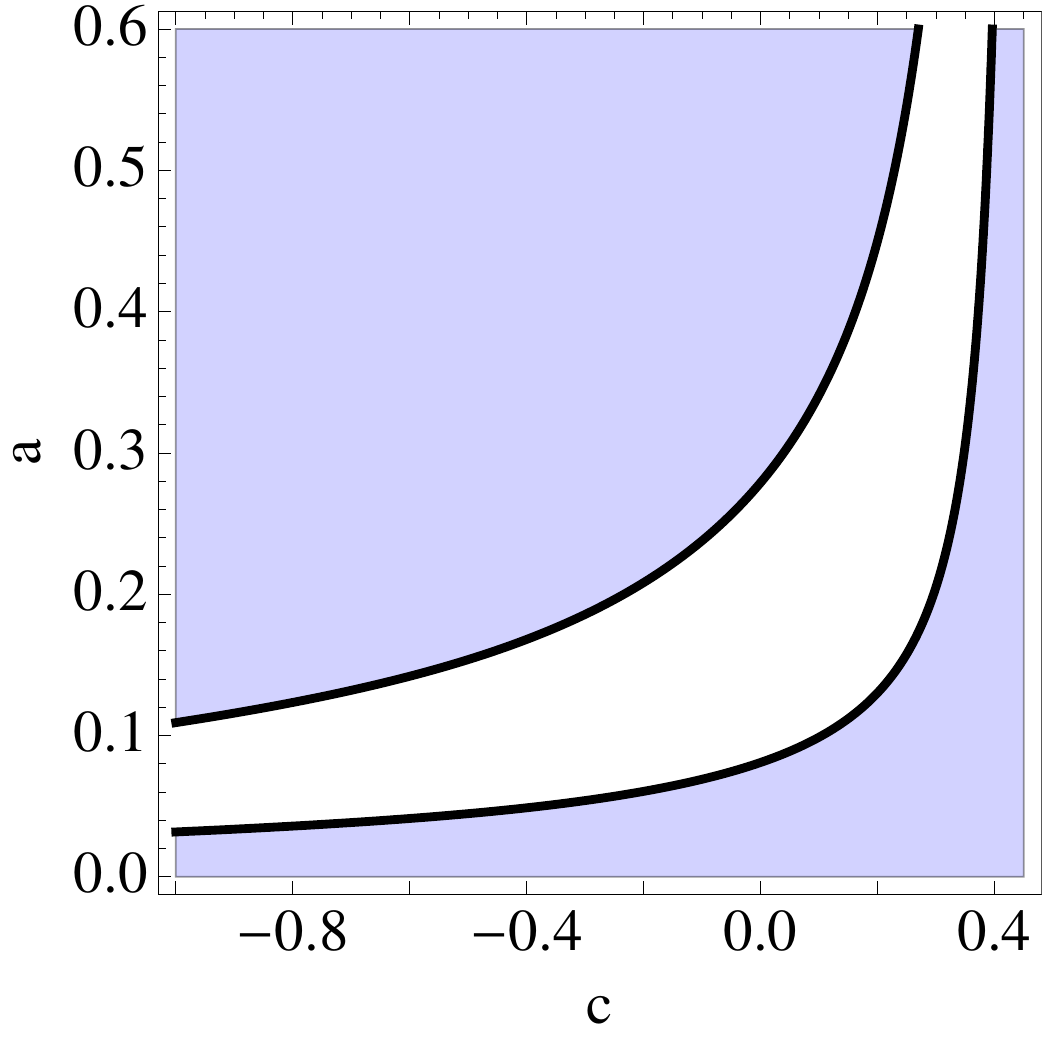} \hfill
\caption{\it Contour plot of $\Delta a_\mu$. We have considered  $c_{\mu_L}=0.4$, $c_{\mu_R}=0.5$.}
\label{fig:Damu}
\end{figure}

Summing up all the contributions we define the total contribution to $\Delta a_\mu$ as
\be
\Delta a_\mu=\Delta a_\mu^C+\Delta a_\mu^H+\Delta a_\mu^{\mathcal N}
\ee
where $\Delta a_\mu^C$ ($\Delta a_\mu^H$) is the contribution from the charged VLL ($\widetilde{\mathcal L},\,\widetilde{\mathcal R}$) and the neutral gauge bosons ($Z,\, Z_n,\,\gamma_n$), and their KK modes (the Higgs boson), and $\Delta a_\mu^{\mathcal N}$ the contribution from the neutral VLL ($\mathcal N$) and the charged gauge boson  and its KK modes ($W,\,W_n$).
In Fig.~\ref{fig:Damu} we show the 95\% CL allowed region in the plane $(c,a)$ which provides the experimental value for the muon AMM $\Delta a_\mu^{\rm exp}$ given by Eq.~(\ref{deviation}).
Notice that, as parametrically~\footnote{Except for the tiny (subleading) contribution from the Higgs boson in the loop which goes as $|U_L^{31}|^2+|U_R^{21}|^2$.} $\Delta a_\mu\propto U_L^{31}U_R^{21}$, the allowed region in the plane $(c,a)$ is entirely determined by the mixing angles between the VLL and the muon, which in turn are determined from the electroweak bounds on the observable $\delta g_{L,R}$ in Eq.~(\ref{bounds}).

We can see that the allowed region is localized towards the IR so that it implies a degree of compositeness for the VLL. In particular extending the results to the limiting region where $a\to 1$ we can see that the allowed region from the muon AMM implies the absolute upper bound $c\lesssim 0.42$. However as wee will see next still there are other experimental and theoretical constraints which restrict the allowed region for VLL to encompass the muon AMM.

%
%


\section{Other constraints}
\label{sec:other}
In this section we will present the main constraints from electroweak observables to VLL. First of all, VLL mix with the muon and thus are subject
to strong constraints in the measurement of the $Z\bar\mu\mu$ coupling, as we already have explained in Sec.~\ref{sec:Zmumu}. Second, the presence of VLL modify the universal (oblique) observables
and thus their contribution can be encoded in their correction to the $S$, $T$ and $U$ observables~\cite{Peskin:1991sw}. This will constraint the allowed region in the $(c,a)$ plane. The corresponding constraints coming 
from the correction to the electroweak observables from Kaluza-Klein modes of gauge bosons and fermions were already summarized in Sec.~\ref{sec:model} and will be taken into account in the present one. Moreover, the presence of VLL running in loops should contribute to the decay rate $\Gamma(H\to\gamma\gamma)$, which has been measured at LHC7 and 8 TeV, and is being measured at LHC13 TeV, and to the Higgs quartic coupling $\beta$-function triggering an instability of the electroweak vacuum faster than in the SM. Both effects, as we will see, will further constraint the allowed region. Finally we will need to take into account present experimental bounds from direct searches at LHC.

\subsection{Oblique corrections}

The relevant Lagrangian in the interaction basis is given by
\begin{align}
\mathcal L &\supset \frac{g}{2}W_\mu^3\left[\bar\nu_L\gamma^\mu\nu_L+\bar N\gamma^\mu N -\bar\mu\gamma^\mu \mu- \overline{\mathcal L}\gamma^\mu\mathcal L
\right]\nonumber\\
-&\frac{g^\prime}{2} B_\mu\left[\bar\nu_L\gamma^\mu\nu_L+\bar\mu \gamma^\mu\mu+\bar N\gamma^\mu N+\overline{\mathcal L}\gamma^\mu \mathcal L+2 \,\overline{\mathcal R}\gamma^\mu\mathcal R \right]\nonumber\\
+&\frac{g}{2}W_\mu^1\left[\bar\nu_L\gamma^\mu \mu_L+\bar N\gamma^\mu \mathcal L+h.c.
\right]\,.
\end{align}
It can be written in the mass eigenstate basis $(\widetilde\mu,\widetilde{\mathcal L},\widetilde{\mathcal R})$ by making the change
\begin{align}
\mu_{L,R}&=U_{L,R}^{11}\widetilde\mu_{L,R}+U_{L,R}^{12}\widetilde{\mathcal L}_{L,R}+U_{L,R}^{13}\widetilde{\mathcal R}_{L,R},\nonumber\\
\mathcal L_{L,R}&=U_{L,R}^{21}\widetilde\mu_{L,R}+U_{L,R}^{22}\widetilde{\mathcal L}_{L,R}+U_{L,R}^{23}\widetilde{\mathcal R}_{L,R},\nonumber\\
\mathcal R_{L,R}&=U_{L,R}^{31}\widetilde\mu_{L,R}+U_{L,R}^{32}\widetilde{\mathcal L}_{L,R}+U_{L,R}^{33}\widetilde{\mathcal R}_{L,R}\,.
\end{align}
After using the expressions for $U_{L,R}$ given in Eq.~(\ref{Aprox2}), and neglecting the matrix elements $U_{L}^{31}$ and $U_R^{21}$ from Eq.~(\ref{Ubounds}), we obtain
\be
\mu\simeq\widetilde \mu,\quad \begin{pmatrix}\mathcal L_{L,R}\\ \mathcal R_{L,R}\end{pmatrix}\simeq \begin{pmatrix} \cos\theta_{L,R}& \sin\theta_{L,R}\\-\sin\theta_{L,R}& \cos\theta_{L,R}\end{pmatrix}\begin{pmatrix} \widetilde{\mathcal L}_{L,R}\\ \widetilde{\mathcal R}_{L,R}\end{pmatrix}
\ee
which can be used to compute Eq.~(\ref{ST}). For the case considered in Sec.~\ref{sec:cLcR}, the contribution of VLL to the $S$ and $T$ observables can be written as~\cite{Ellis:2014dza}
\begin{align}
\Delta S&=8\pi\left[\Pi^\prime(M)-\frac{3}{4}\left(\Pi^\prime(M_{\widetilde{ \mathcal L}})+\Pi^\prime(M_{\widetilde {\mathcal R}})\right)+\frac{1}{2}\Pi^\prime(M_{\widetilde{\mathcal L}},\,M_{\widetilde{\mathcal R}})  
\right]\nonumber\\
\Delta T&=\frac{2\pi}{s_W^2 m_W^2}\left[  \Pi(M,\,M_{\widetilde{\mathcal L}}) +\Pi(M,\,M_{\widetilde{\mathcal R}})-\frac{1}{2} \Pi(M_{\widetilde{\mathcal L}},\,M_{\widetilde{\mathcal R}})
\right]
\end{align}
where the self-energies from fermions with masses $m_a$ and $m_b$ propagating in the loop, $\Pi(p^2;m_a,m_b)$ and $d\Pi(p^2;m_a,m_b)/dp^2$ are defined at $p^2=0$, as $\Pi(0;m_a,m_b)\equiv \Pi (m_a,m_b)$ and $\left. d\Pi(p^2;m_a,m_b)/dp^2\right|_{p^2=0}\equiv \Pi^\prime(m_a,m_b)$, with
\begin{align}
\Pi(m_a,m_b)&=\frac{1}{32\pi^2}\frac{1}{(m_a^2-m_b^2)}
\left[m_a^4-m_b^4-2 m_a^4 \log m_a^2+2m_b^4\log m_b^2\right.\nonumber\\
&-\left. 4m_a m_b(m_a^2-m_b^2-m_a^2\log m_a^2+m_b^2\log m_b^2)\right]\nonumber\\
\Pi^\prime(m_a,m_b)&=\frac{1}{144\pi^2}\frac{1}{(m_a^2-m_b^2)^3}\left[-2 m_a^6+2 m_b^6+18 m_a^2 m_b^2(m_a^2-m_b^2)\right.
\nonumber\\
+& 6m_a^4(m_a^2-3m_b^2)\log m_a^2-6m_b^4(m_b^2-3 m_a^2) \log m_b^2\nonumber\\
-&\left. 9 m_a m_b(m_a^4-m_b^4-2 m_a^2 m_b^2 \log(m_a^2/m_b^2))
\right]
\end{align}
and where $\Pi^\prime(m_b)=\lim_{m_a\to m_b}\Pi^\prime(m_a,m_b)$.

\begin{figure}[htb]
\centering
\includegraphics[width=7cm]{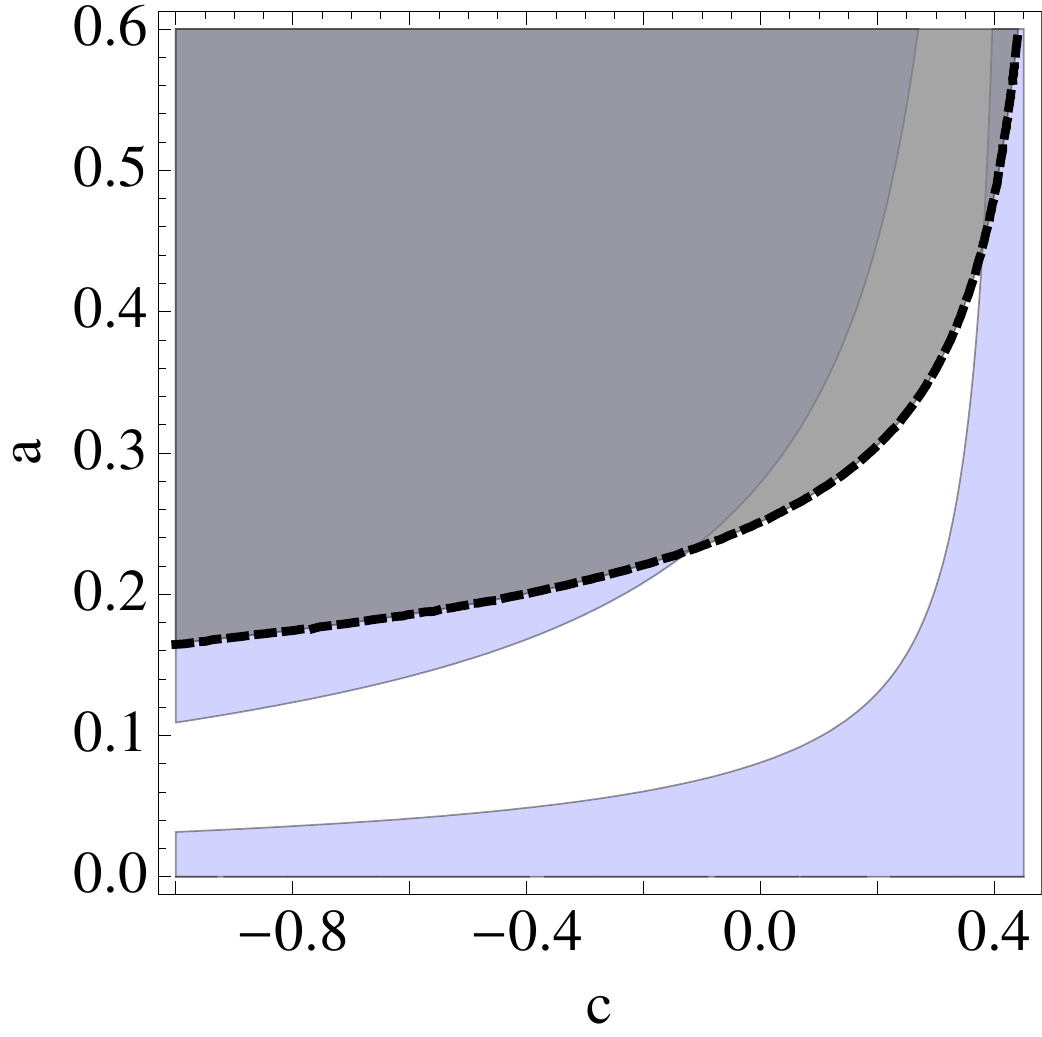} \hfill
\includegraphics[width=7cm]{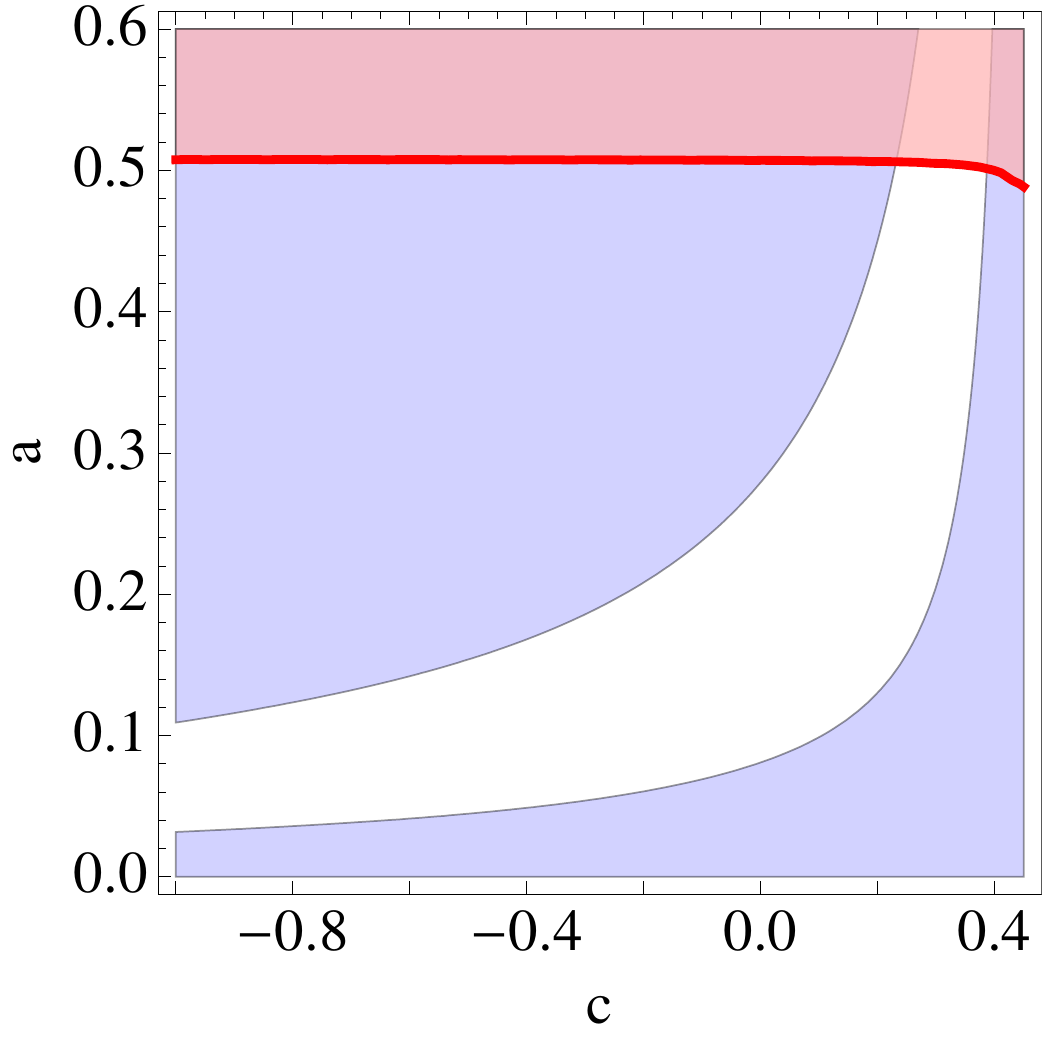} 
\caption{\it Bounds imposed by oblique observables (left panel) and by $H\to\gamma\gamma$ (right panel). The region allowed by the muon AMM is superimposed.}
\label{fig:STgamma}
\end{figure} 
We will show our results in the plane $(c,a)$. The region allowed by oblique parameters consistent with the experimental data~\cite{Agashe:2016kda}, Eq.~(\ref{eq:ST}), at 95\% CL, is given in the left panel of Fig.~\ref{fig:STgamma}, where the excluded region is shadowed, and we have superimposed (as we will do in all the plots from here on) the region allowed by the muon AMM. We can see that the universal (oblique) observables impose upper bounds on the parameters $a$ and $c$, as $a\lesssim 0.42$ and $c\lesssim 0.39$. This condition imposes the mild bound on the mass of the lightest eigenstate as
$M_{\widetilde{\mathcal L}}\gtrsim 230$ GeV.

\subsection{$H\to \gamma\gamma$ } 
The interactions of the Higgs with the VLL (\ref{Y4D2}) generate, when the charged VLL propagate in the loop, an extra contribution to the processes $H\to \gamma\gamma$.
Taking only into account the contribution of the $W$ boson, the top quark $t$ and other extra fermions $f$, and  
neglecting the off-diagonal elements in (\ref{Y4D2}), we can 
write~\cite{Carena:2012xa}
\begin{align}
\varGamma(H\to \gamma\gamma)=\frac{G_F\alpha^2m^3_h}{128 \sqrt{2}\pi^3} \bigg| A_{1}(\tau_W)+N_c Q^2_tA_{1/2}(\tau_t)+
\sum_{f=\widetilde{\mathcal L},\widetilde {\mathcal R}}\frac{v Y_{f f}}{M_f} A_{1/2}(\tau_f) \bigg|^2
 \end{align}
where   $N_c = 3$ is the number of colors, $Q_t = +2/3$ is the top quark
electric charge in units of $|e|$, $\tau_i\equiv4m^2_i/m^2_H$, $i=W,t,f$,  and 
\begin{align}
A_{1}(x)=&-x^2[2x^{-2}+3x^{-1}+3(2x^{-1}-1)f(x^{-1})] \,,\\
A_{1/2}(x)=&2x^2[x^{-1}+(x^{-1}-1)f(x^{-1})]
  \end{align}
with
  \begin{align}
  f(x)&=\left\{ \begin{array}{cc}  
                 \arcsin^{2}(\sqrt{x}),& 0<x<1\\
                 -\ln^{2}(\sqrt{x}+\sqrt{x-1})+\frac{1}{4}\pi^{2}+i\pi\ln\left(\sqrt{x}+\sqrt{x-1}\right),& x>1.
              \end{array}\right.
\end{align}

The observable measured by the ATLAS and CMS Collaborations at LHC is the Higgs signal strength $\widehat\mu$ defined as
\begin{align}
\widehat\mu&=\frac{\left.\sigma(pp\to H) \cdot BR(H\to\gamma\gamma\right)|_{\rm obs}}{\left.\sigma(pp\to H) \cdot BR(H\to\gamma\gamma\right)|_{\rm SM}}
 \end{align}
with a combined value for ATLAS and CMS given by $\widehat\mu=1.09\pm 0.11$~\cite{Khachatryan:2016vau}.

The contribution of the charged VLL, $\widetilde{\mathcal L}$ and $\widetilde{\mathcal R}$, to $\widehat\mu$ is positive and its present experimental value already excludes a region in the plane $(c,a)$ as it is shown in the right panel of Fig.~\ref{fig:STgamma}. As it is clear from Fig.~\ref{fig:STgamma}, the region excluded by $\widehat \mu$, at 95\% CL, is inside the region already excluded by oblique observables and does not restrict further the region allowed by the muon AMM. However future measurements of the Higgs strength $\widehat\mu$ could possibly exclude additional regions in the plane $(c,a)$. For instance a hypothetical (much stronger) bound as $\widehat\mu<1.01$ would translate into the upper bounds $c\lesssim 0.15$ and $a\lesssim 0.11$ which translate into the lower bound on the mass of the lightest VLL, $M_{\widetilde{\mathcal L}}\gtrsim 800$~GeV.

\subsection{The stability of the electroweak minimum}

An important (theoretical) constraint is the (in)stability of the electroweak minimum for scales larger than the mass of VLL, and thus much larger than the electroweak scale. For large values of the Higgs field $H$, the tree-level Higgs potential can be approximated by 
\be
V_0(H)\simeq \lambda(\mu) |H|^4
\ee
where $\mu\simeq |H|$. This effect already appears in the SM due to the contribution of the top quark to the renormalization group equations (RGE) of the Higgs quartic coupling $\lambda$ in the 4D theory.

It is well known that in the SM, and for the measured values of the top quark and Higgs boson masses, the electroweak vacuum becomes unstable (i.e.~$\lambda<0$) at a scale $\mu_I\simeq 10^{10}$ GeV, although the tunneling lifetime from the electroweak vacuum to the false vacuum is much larger than the age of the universe~\cite{Casas:1994qy,Degrassi:2012ry}.

\begin{figure}[htb]
\centering
\includegraphics[width=7.55cm]{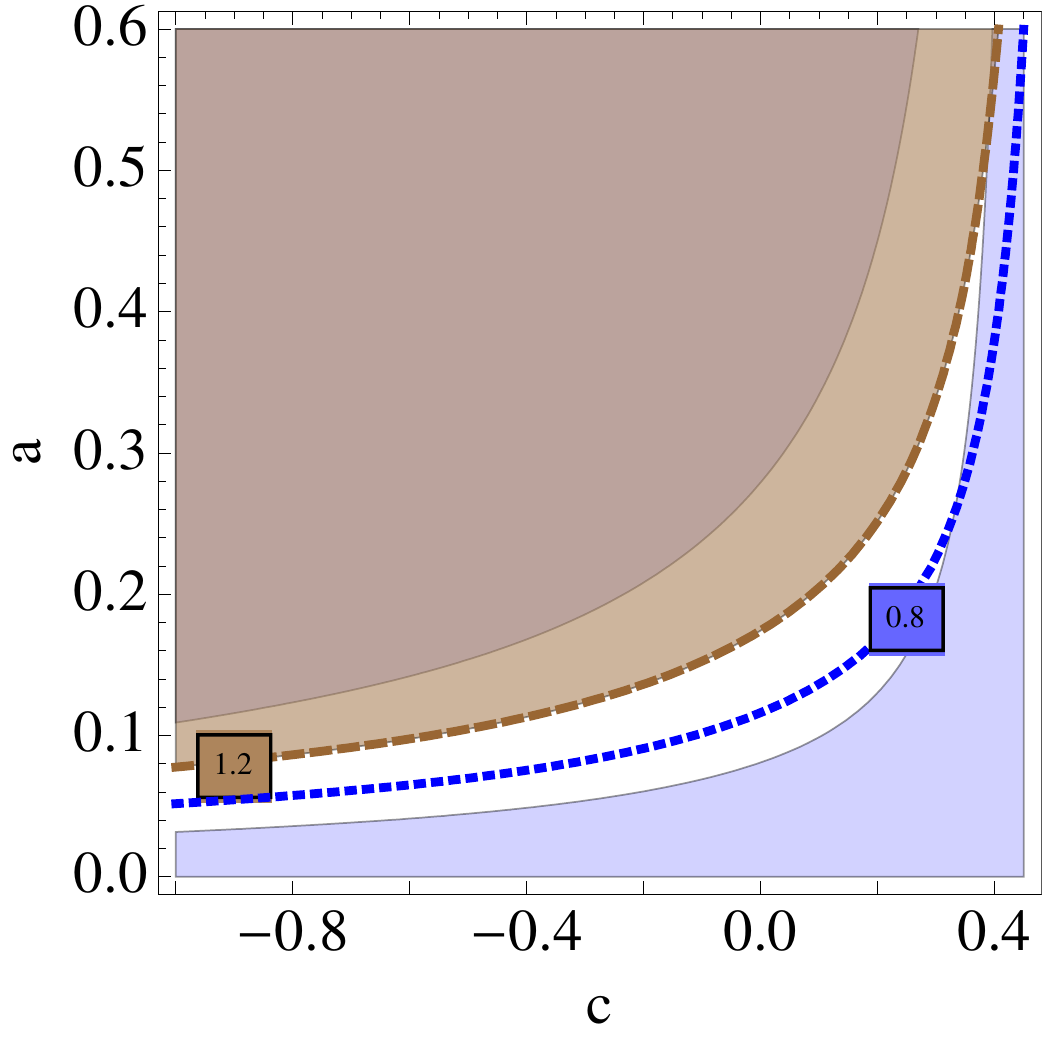} \hfill
\includegraphics[width=7.55cm]{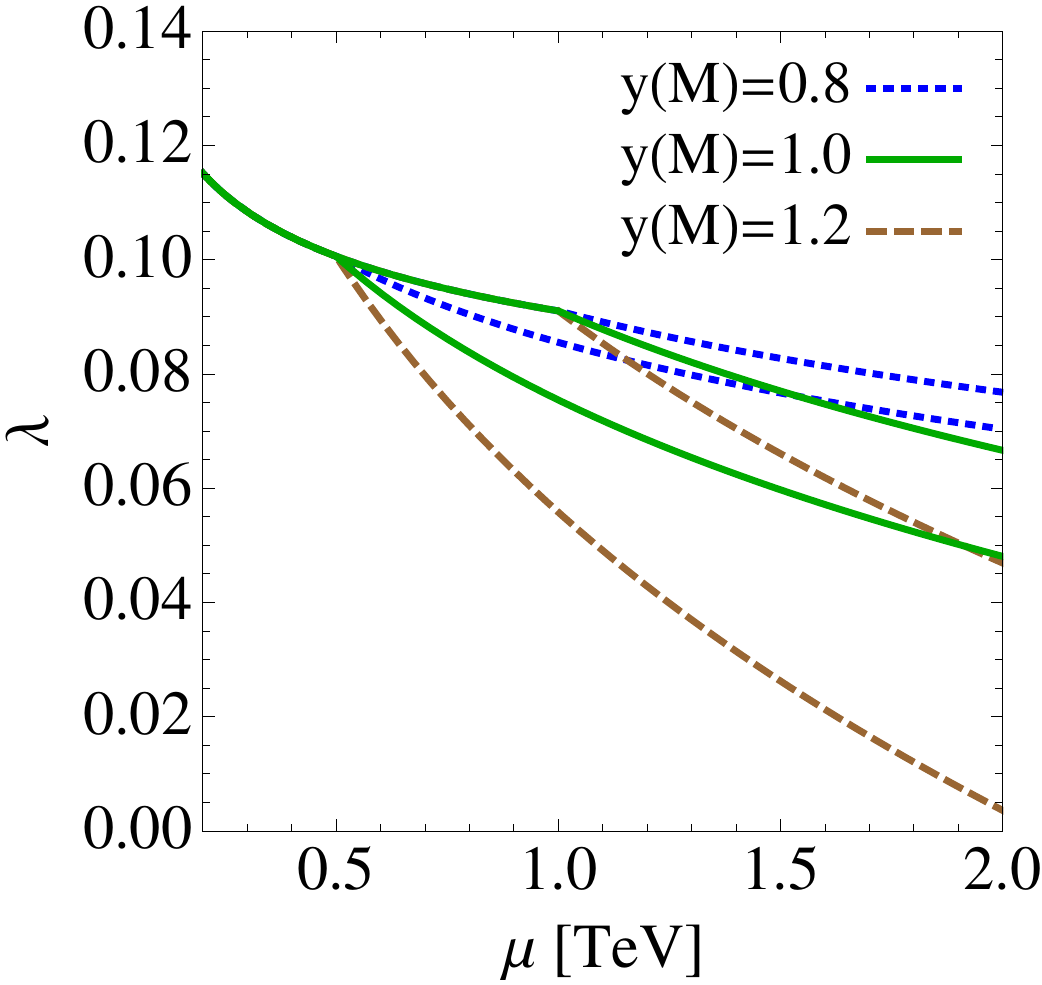}
\caption{\it Left panel: Contour plot of $y(c,a)$ in the plane $(c,a)$. Right panel: RGE evolution of $\lambda$ for $y(M)=0.8$ (dotted blue lines), $y(M)=1$ (solid red lines) and $y(M)=1.2$ (dashed green lines). For every value of $y$, $M=1$ TeV (upper line) and $M=0.5$ TeV (lower line).}
\label{fig:y1}
\end{figure} 
In the presence of VLL with Yukawa couplings $\left.y(\mu)\right|_{\mu=M}\equiv a M(c)/v$ to the Higgs field, see Eq.~(\ref{Y4D2}), the instability problem is more acute as the quartic coupling is driven faster to negative values. In fact VLL contribute to the SM RGE by~\cite{ArkaniHamed:2012kq,Joglekar:2012vc}
\be
\Delta\beta_\lambda=\frac{1}{16\pi^2}(8\lambda y^2-4 y^4), \  \Delta\beta_{h_t}=\frac{1}{16\pi^2}2 h_t y^2
\label{deltarge}
\ee
and for large values of the Yukawa coupling $y$, the (quartic) term $y^4$ in Eq.~(\ref{deltarge}) drives rapidly $\lambda$ to negative values. Particles which are (almost) localized on the TeV brane, such as the Higgs, the top quark or the VLL, only contribute to the running above their mass and below the energy $\mu=m_{KK}$,  where the theory is 4D~\cite{Randall:2001gc}. For scales $\mu>m_{KK}$ they contribute like the bulk fields which represent their preonic constituents. From the holographic point of view this is due to the fact that TeV brane fields are the bound states of the near conformal field theory (CFT) at higher energy scales. In fact the running for $\mu>m_{KK}$ depends on the  particular preonic constituents and it is thus very much model dependent. In this paper we will just present the 4D running of the Higgs quartic coupling, see the right panel of Fig.~\ref{fig:y1}. 
The 5D running, and thus the full problem of the stability of the electroweak minimum, is model dependent and beyond the scope of the present paper, and it is postponed for a future work. We plot the running of $\lambda$ for different values of the coupling $y(\mu)$ at the scale $\mu=M$, $y(M)=0.8,1,1.2$ and for different values of $M$, $M=0.5,1$ TeV. For scales $\mu<M$, VLL are decoupled and the running is purely the SM one. For scales $\mu>M$, VLL are integrated in and they contribute to $\beta_\lambda$ triggering a quick descent of $\lambda$. As VLL are active only for $\mu>M$, the smaller the value of $M$, the faster $\lambda$ goes to zero. In fact we can see that for $M=0.5$ TeV the value of $\lambda(m_{KK})$ gets very close to zero for $y(M)=1.2$, which puts an absolute upper bound on $y(M)$ as $y(M)\lesssim 1.2$. This bound translates into the upper bounds $c\lesssim 0.37$ and $a\lesssim 0.36$, i.e.~a lower bound on the mass of the lightest VLL as $M_{\widetilde{\mathcal L}}\gtrsim 270$ GeV.
For larger values of $M$ and/or smaller values of $y(M)$, $\lambda(m_{KK})>0$ and the theory is safe from the 4D point of view. Contour plots of $y(M)$ are shown in the left plot of Fig.~\ref{fig:y1}.

\subsection{Collider phenomenology}

Heavy leptons can be produced in pairs at lepton colliders and by Drell-Yan processes at hadron colliders, and in particular at the LHC, with cross-sections $\sigma(pp\to Z^*/\gamma^*\to \widetilde{\mathcal L}^+\widetilde{\mathcal L}^-)$ which depend on the center of mass energy, the mass and the couplings of VLL to $Z/\gamma$. In our model VLL couple to electroweak gauge bosons with SM couplings.
In particular VLL could have been produced at LEP2 in the process $e^+ e^-\to Z/\gamma\to \widetilde{\mathcal L}^+ \widetilde{\mathcal L}^-$ settling the lower bound $M_{ \widetilde{\mathcal L}}>101.2$ GeV~\cite{Achard:2001qw}. More recently
a search for heavy leptons decaying into $Z$ and muons is done by the ATLAS collaboration~\cite{Aad:2015dha} based on $pp$ collision data taken at $\sqrt{s}=8$ TeV with an integrated luminosity of 20.3 fb$^{-1}$. VLL are excluded at 95\% CL for masses $M_{ \widetilde{\mathcal L}}<168$ GeV. As we will see this relevant region is already excluded by the other constraints and after imposing that the correct value of the muon anomalous magnetic moment is reproduced.

Stronger bounds are expected in the future based on collisions at $\sqrt{s}=13$ TeV although the production cross-section decreases very fast for larger values of the masses of the VLL. For instance it turns out that $\sigma(pp\to Z/\gamma^*\to \widetilde{\mathcal L}^+\widetilde{\mathcal L}^-)\lesssim \mathcal O(1)$ fb for $M_{ \widetilde{\mathcal L}}\gtrsim 500$~GeV~\cite{ArkaniHamed:2012kq}, and so a full-fledged collider study should be done to put bounds on $M_{ \widetilde{\mathcal L}}$ based on $\sqrt{s}=13$ TeV data.

Once the lightest VLL, $\widetilde{\mathcal L}$, is produced it decays through the channels $\widetilde{\mathcal L}\to \widetilde\mu Z, \nu_L W, \tilde\mu H$. The relevant couplings are given by the Lagrangian
\begin{align}
\mathcal L_{VLL}=&g^Z_{\widetilde{\mathcal L}_{L,R}} Z_\mu \overline{\widetilde\mu}_{L,R}\gamma^\mu \widetilde{\mathcal L}_{L,R}+g^W_{\widetilde{\mathcal L}_{L}}W_\mu \overline{\nu}_L \gamma^\mu \widetilde{\mathcal L}_L\nonumber\\
+&\overline{\widetilde \mu}H\left(
\frac{Y_{12}+Y_{21}}{2\sqrt{2}}-\gamma_5 \frac{Y_{12}-Y_{21}}{2\sqrt{2}}\right)\widetilde{\mathcal L}
+h.c.
\end{align}
where the matrix $Y$ is defined in Eq.~(\ref{Y4D2}) and
\begin{align}
g^Z_{\widetilde{\mathcal L}_L}&=\frac{U_L^{31}}{\sqrt{2}}(g^Z_{\mu_L}-g^Z_{\mu_R})=-U_L^{31} \frac{g}{2\sqrt{2} c_W},\nonumber\\
g^Z_{\widetilde{\mathcal L}_R}&=\frac{U_R^{21}}{\sqrt{2}}(g^Z_{\mu_L}-g^Z_{\mu_R})=-U_R^{21} \frac{g}{2\sqrt{2} c_W}\nonumber\\
g^W_{\widetilde{\mathcal L}_L}&=\frac{g(1+a)}{2}U_L^{31},\quad \frac{Y_{12}\pm Y_{21}}{2\sqrt{2}}=\frac{(1+a)M}{4v}\left(U_L^{31}\mp U_R^{21}  \right)
\label{acoplosdecay}
\end{align}
Two observations from Eq.~(\ref{acoplosdecay}) are now in order
\begin{itemize}
\item
As, from Eq.~(\ref{bounds}), $|U_L^{31}|,\, |U_R^{21}|\lesssim 0.02$ the gauge couplings are tiny. In particular the gauge couplings with the $Z$ and the $W$ are $\lesssim 6\times 10^{-3}$.
\item
 The Yukawa coupling remains perturbative in the whole region where $M\lesssim 2.5$ TeV, for which $|Y_{12}/(2\sqrt{2})|,\,|Y_{21}/(2\sqrt{2})|\lesssim 0.07$. In the opposite extreme, for light VLL, say $M\gtrsim 250$ GeV, we find $|Y_{12}/(2\sqrt{2})|,\,|Y_{21}/(2\sqrt{2})|\gtrsim 0.007$. 
\end{itemize}

The decay width for the channel $\widetilde{\mathcal L}\to \widetilde\mu H$ is given by%
\begin{align}
\Gamma(\widetilde{\mathcal L}\to \widetilde\mu H) &= \frac{(|Y_{12}|^2+|Y_{21}|^2)}{64\pi} M_{\widetilde{\mathcal L}}\left(1-\frac{m_H^2}{M_{\widetilde{\mathcal L}}^2}  \right)^2 \nonumber \\
&=\frac{g^2(1-a^2)(1+a)}{256\pi}\,\frac{M^3}{m_W^2} \left( |U_L^{31}|^2 + |U_R^{21}|^2 \right) \left(1-\frac{m_H^2}{(1-a)^2 M^2}  \right)^2 ,
\label{H}
\end{align}
while the decay widths for the channels  $\widetilde{\mathcal L}\to \widetilde\mu Z$ and  $\widetilde{\mathcal L}_L\to \nu_L W$ are given by
\begin{align}\label{W}
\Gamma(\widetilde{\mathcal L}_L\to \nu_L W)=&\frac{\left( g^W_{\widetilde{\mathcal L}_L}\right)^2}{32\pi}\frac{M_{\widetilde{\mathcal L}}^3}{m_W^2}\left( 1-\frac{m_W^2}{M_{\widetilde{\mathcal L}}^2}  \right)^2 \left( 1+2 \frac{m_W^2}{M_{\widetilde{\mathcal L}}^2}  \right)\\
=& \frac{g^2 (1-a^2)^2(1-a)}{128\pi}\, |U_L^{31}|^2\,\frac{M^3}{m_W^2}\left( 1-\frac{m_W^2}{(1-a)^2M^2}  \right)^2 \left( 1+2 \frac{m_W^2}{(1-a)^2M^2}  \right)\nonumber
\end{align}
\begin{align}\label{Z}
\Gamma(\widetilde{\mathcal L}\to \widetilde\mu Z)&=\frac{\left( g^Z_{\widetilde{\mathcal L}_L}\right)^2+\left( g^Z_{\widetilde{\mathcal L}_R}\right)^2}{32\pi}\frac{M_{\widetilde{\mathcal L}}^3}{m_Z^2}\left( 1-\frac{m_Z^2}{M_{\widetilde{\mathcal L}}^2}  \right)^2 \left( 1+2 \frac{m_Z^2}{M_{\widetilde{\mathcal L}}^2}  \right)\\
=& \frac{g^2 (1-a)^3}{256\pi}\,\left( |U_L^{31}|^2+ |U_R^{21}|^2\right)\,\frac{M^3}{m_W^2}
\left( 1-\frac{m_Z^2}{(1-a)^2M^2}  \right)^2 \left( 1+2 \frac{m_Z^2}{(1-a)^2M^2}  \right)\nonumber
\end{align}

\begin{figure}[htb]
\centering
\includegraphics[width=7.55cm]{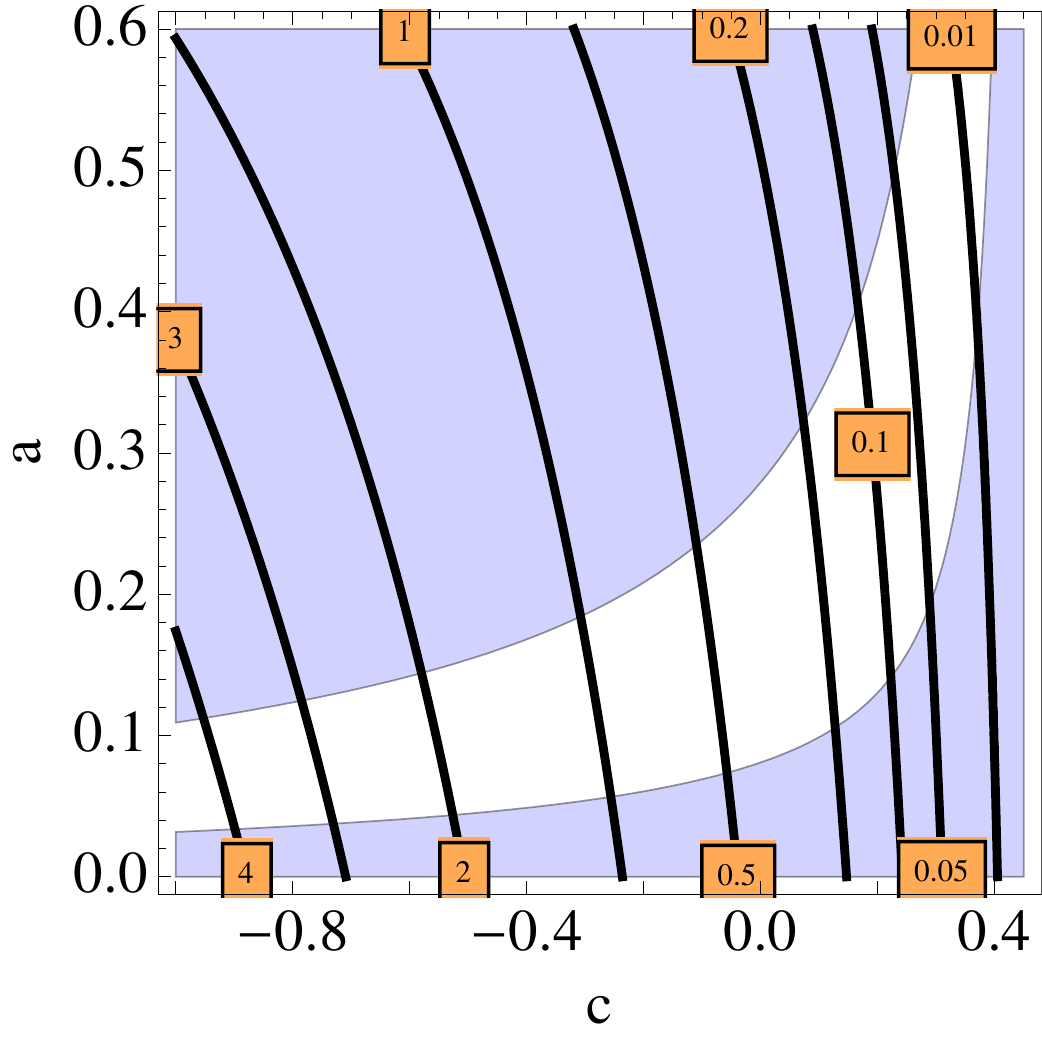} \hfill
\includegraphics[width=7.55cm]{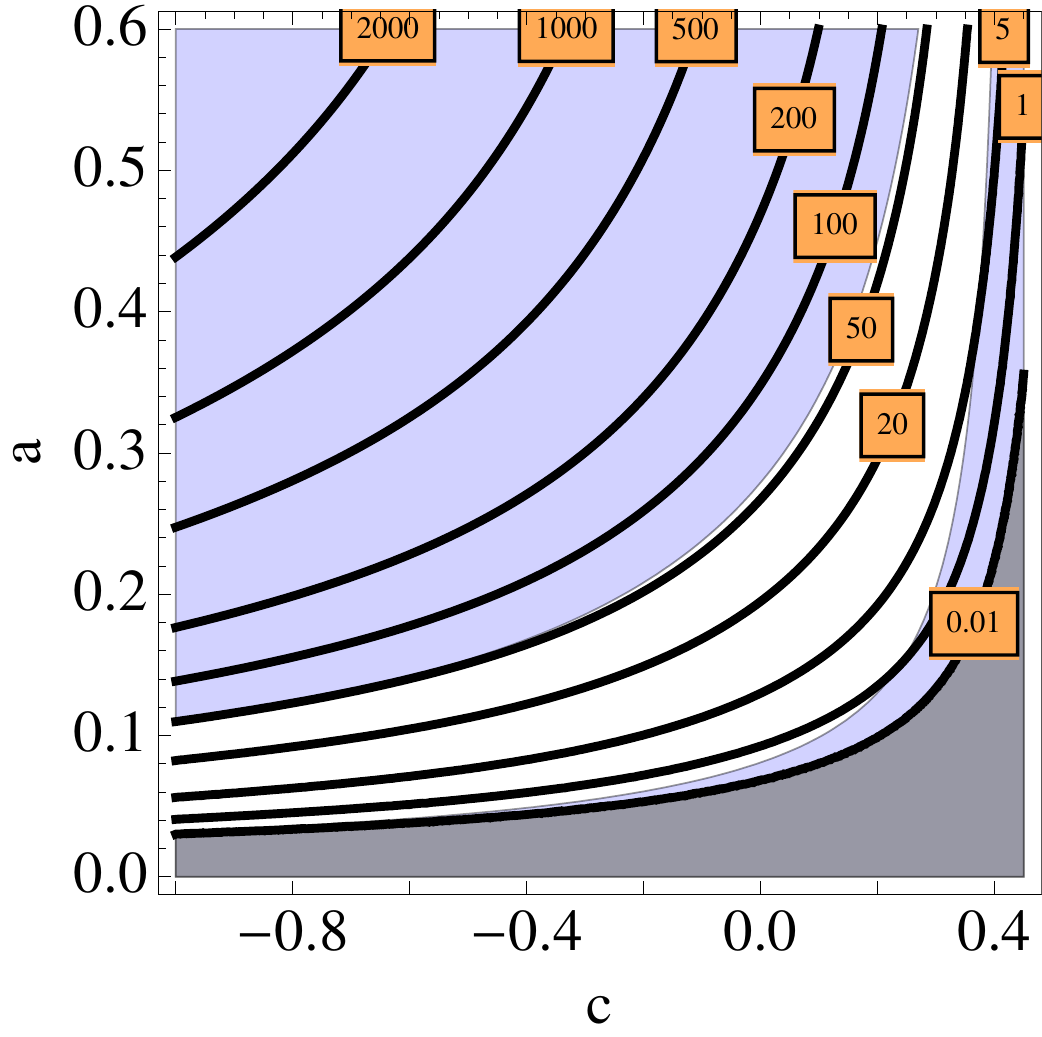}
\caption{\it Left panel: Contour plot of $\Gamma_{\widetilde{\mathcal L}}$ in GeV. Right panel: Contour plot of $\Gamma_{\mathcal N}$ in GeV. The black shaded area in the right panel is the two-body excluded region $a < m_W / M$, corresponding to the three-body decay channel.}
\label{fig:mfp}
\end{figure}

The total width of the lightest VLL is given by $\Gamma_{\widetilde{\mathcal L}}=\Gamma(\widetilde{\mathcal L}\to \widetilde\mu H)+\Gamma(\widetilde{\mathcal L}_L\to \nu_L W)+\Gamma(\widetilde{\mathcal L}\to \widetilde\mu Z)$. We show in the left panel of Fig.~\ref{fig:mfp}  the plot of $\Gamma_{\widetilde{\mathcal L}}$ in the plane $(c,a)$. Its mean free path is given by $c\tau = [1.97\,/\Gamma_{\widetilde{\mathcal L}}(GeV)]  \times10^{-10} \mu m$ so that in all cases the decay is extremely prompt,
\begin{figure}[htb]
\centering
\includegraphics[width=5cm]{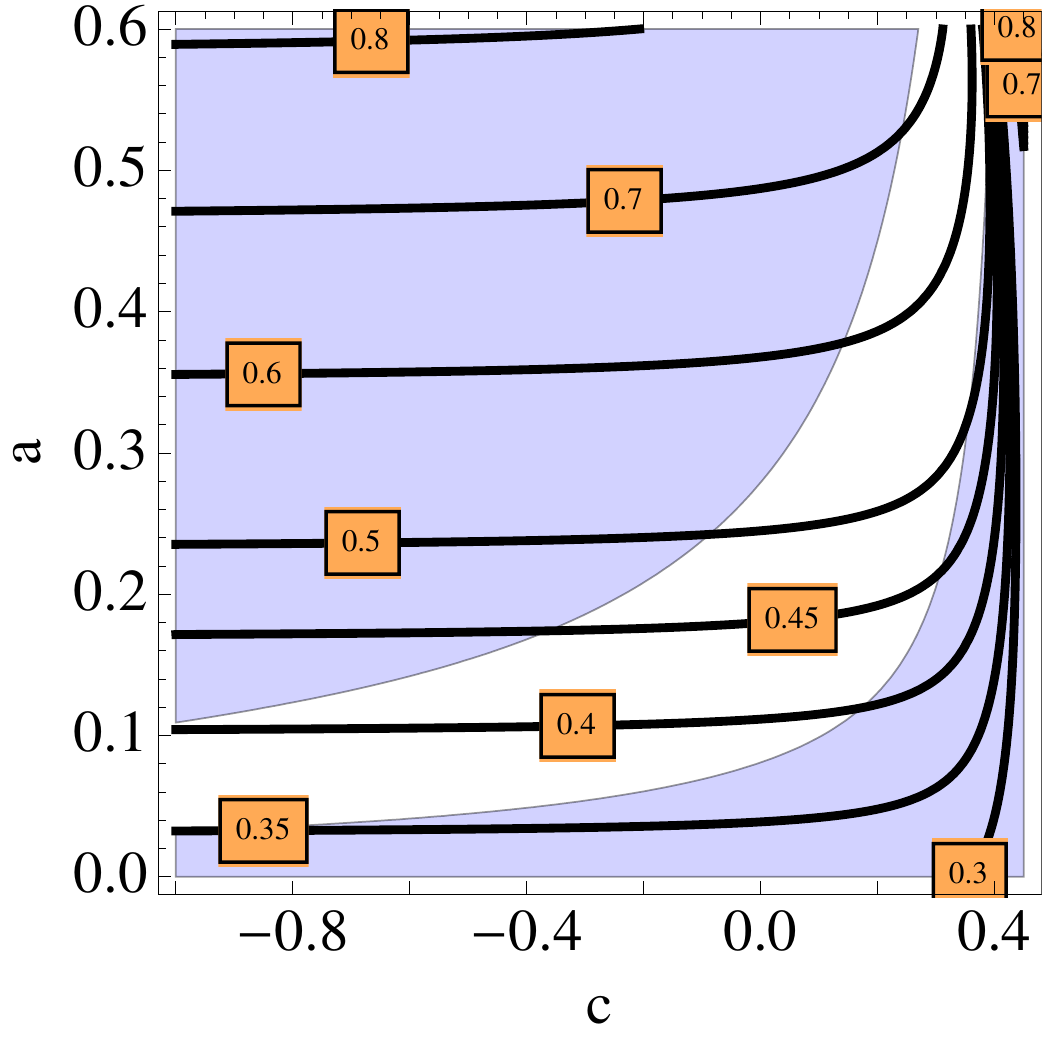} \hfill
\includegraphics[width=5cm]{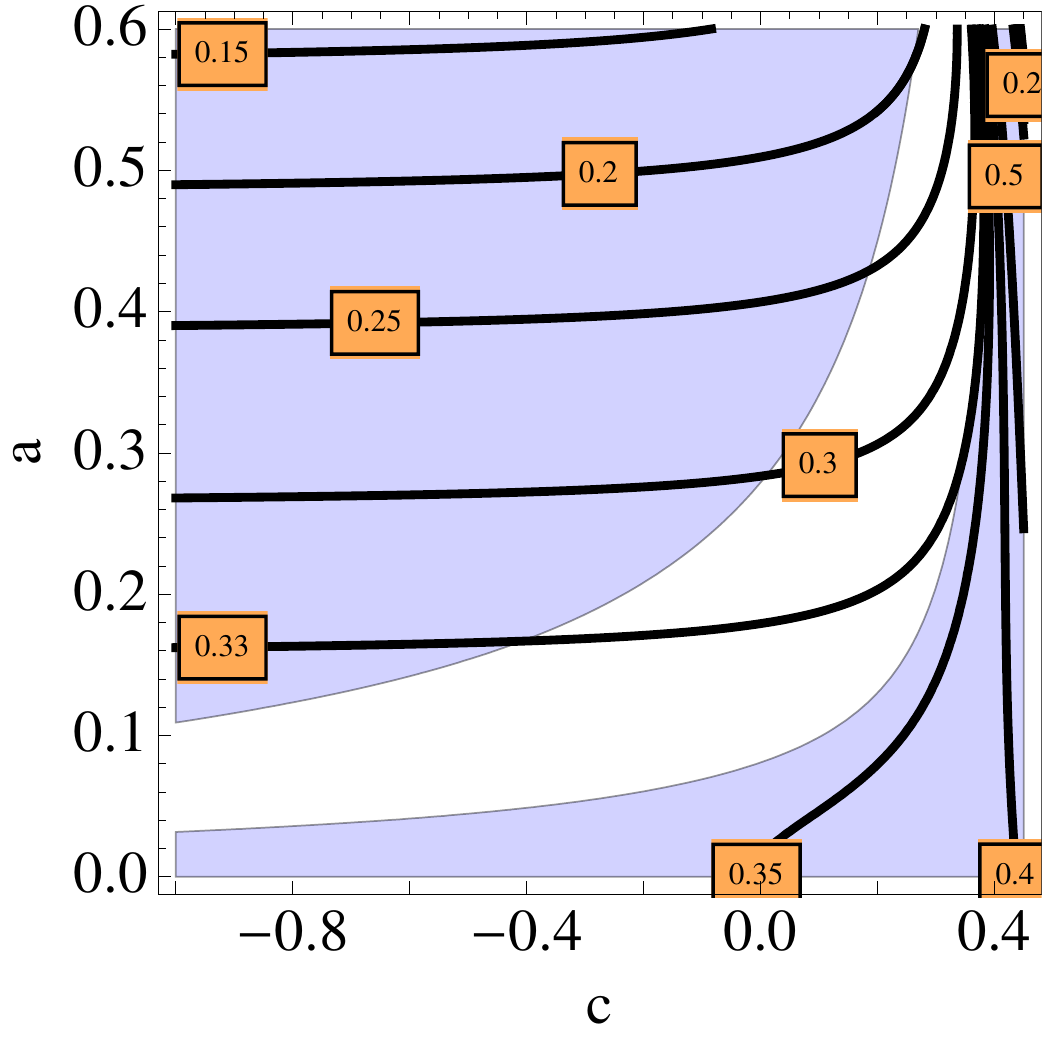} \hfill
\includegraphics[width=5cm]{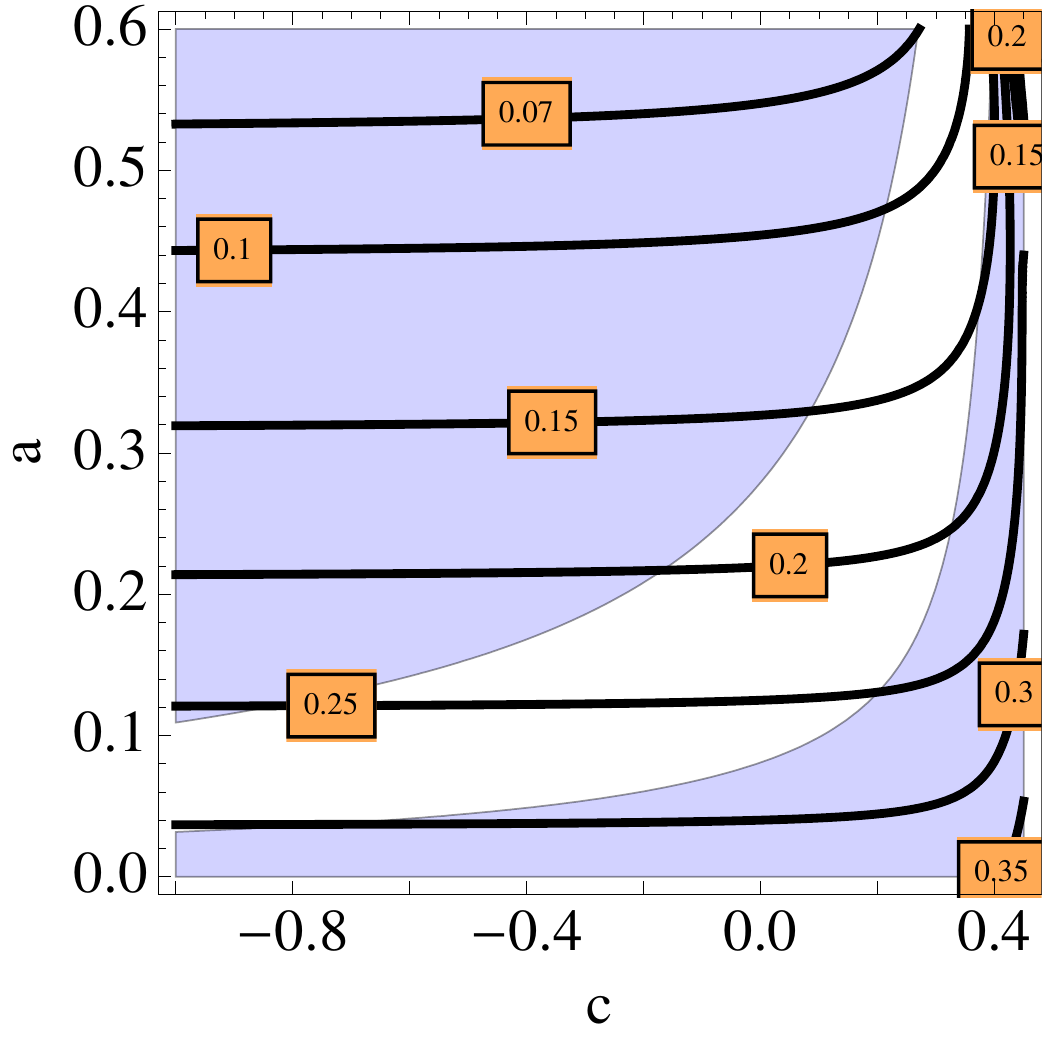}
\caption{\it Branching ratios of the decays  $\widetilde{\mathcal L}\to \widetilde\mu H$ (left panel), $\widetilde{\mathcal L}_L\to \nu_L W$ (middle panel)  and $\widetilde{\mathcal L}\to \widetilde\mu Z$ (right panel).}
\label{fig:Br}
\end{figure} 
as the distance of the secondary decay vertex can never be resolved from the interaction point~\footnote{The distance between the displaced vertices and the interaction point which can be resolved inside the detector is typically given by $c\tau\simeq(75-100)\,\mu m$. For displaced vertices such that $c\tau\lesssim 75\,\mu m$ the particle is called \textit{prompt}. For displaced vertices such that $100\mu m<c\tau<(1-3)\,m$ the particle decays inside the detector and the displaced vertex can be reconstructed. For decay distances $c\tau>3m$ the particle decays outside the detector and it is called \textit{long-lived}. There are strong constraints on the mass of long-lived charged particles~\cite{Khachatryan:2016sfv} which do not apply to our model.}.
From the partial expressions of the decay rates of $\widetilde{\mathcal L}$ into the different channels in Eqs.~(\ref{H}), (\ref{W}) and (\ref{Z}) we can decompose the total rate as $\Gamma_{\widetilde{\mathcal L}}=\Gamma_{\widetilde{\mathcal L}_L}+\Gamma_{\widetilde{\mathcal L}_R}$ where $\Gamma_{\widetilde{\mathcal L}_L}$ ($\Gamma_{\widetilde{\mathcal L}_R}$) is the term of $\Gamma_{\widetilde{\mathcal L}}$ proportional to $|U_L^{31}|^2$ ($|U_R^{21}|^2$). As we can see the ratio of contributions to $H:W:Z$ in  $\Gamma_{\widetilde{\mathcal L}_L}$, in the limit of large values of $M$, is equal to $1:2:1$ in agreement with the Goldstone Boson Equivalence Theorem. The same happens for $\Gamma_{\widetilde{\mathcal L}_R}$, except that the $W$ channel does not exist in $\Gamma_{\widetilde{\mathcal L}_R}$, as we are assuming only left-handed neutrinos in doublets. In Fig.~\ref{fig:Br} we show contour lines of the branching ratios corresponding to the different channels $\widetilde{\mathcal L}\to \widetilde\mu H$ (left panel), $\widetilde{\mathcal L}_L\to \nu_L W$ (middle panel)  and $\widetilde{\mathcal L}\to \widetilde\mu Z$ (right panel). We see that in spite of the fact that gauge couplings are much smaller than the Yukawa couplings all the different branching ratios are of the same order of magnitude in most of the parameter space.

The next-to-lightest VLL is the (neutral) vector-like neutrino (VLN) $\mathcal N$ with a mass $M$. VLN are pair produced by Drell-Yan processes at the LHC via a $Z$ gauge boson, $\sigma(pp\to Z\to \mathcal N\mathcal N)$. When $a>m_W/M$ it decays into the channel $\mathcal N\to \widetilde{\mathcal L}\, W$. The region $a>m_W/M$ is shown in the right panel of Fig.~\ref{fig:mfp} from where we can see that it overlaps with the allowed region from all previous constraints~\footnote{In the region $a<m_W/M$ the vector-like neutrino $\mathcal N$ decays through $\mathcal N\to \widetilde{\mathcal L}\, W^*\to \widetilde{\mathcal L} f_1 f_2$ in a three-body decay channel.}. The relevant Lagrangian is
\be
\mathcal L_{VLN}=\frac{g}{2}W^\mu \overline{\mathcal N} \gamma_\mu  \widetilde{\mathcal L}
\ee
and the decay width is given by
\begin{align}
\Gamma(\mathcal N\to \widetilde{\mathcal L}\, W)&=\frac{g^2}{64\pi}\,\frac{M^3}{m_W^2}
\left\{(2-a)^2a^2+[1+(1-a)^2-6(1-a)]\frac{m_W^2}{M^2}-2\frac{m_W^4}{M^4}
\right\}\beta(M)\nonumber\\
\beta(M)&=
\sqrt{\left(a^2-\frac{m_W^2}{M^2} \right)\left((2-a)^2-\frac{m_W^2}{M^2} \right)}
\end{align}
The contour plot of $\Gamma_{\mathcal N}\simeq\Gamma(\mathcal N\to \widetilde{\mathcal L}\, W)$ is shown in the right panel of Fig.~\ref{fig:mfp}. We can see that its decay is also prompt.

\section{Conclusions}
\label{sec:conclusions}

In this paper we have assessed the capability of theories, solving the hierarchy problem by mean of a warped extra dimension, to solve some of the flavor anomalies which appear in the muon sector: in particular the $B\to K^*\mu^+\mu^-$ LHCb anomaly and the anomalous magnetic moment of the muon $a_\mu$. To do that we have considered a particular geometry in the warped dimension where the AdS$_5$ symmetry is strongly perturbed near the IR brane, with a naked singularity in the gravitational metric outside the physical interval (the so-called soft-wall metric). These models, where the minimal 5D SM propagates in the bulk of the extra dimension, have the advantage that, even in the absence of an extra gauge custodial symmetry in the bulk, the contribution to the electroweak observables is strongly suppressed for low values of the mass of gauge KK modes.

\begin{figure}[htb]
\centering
\includegraphics[width=8.5cm]{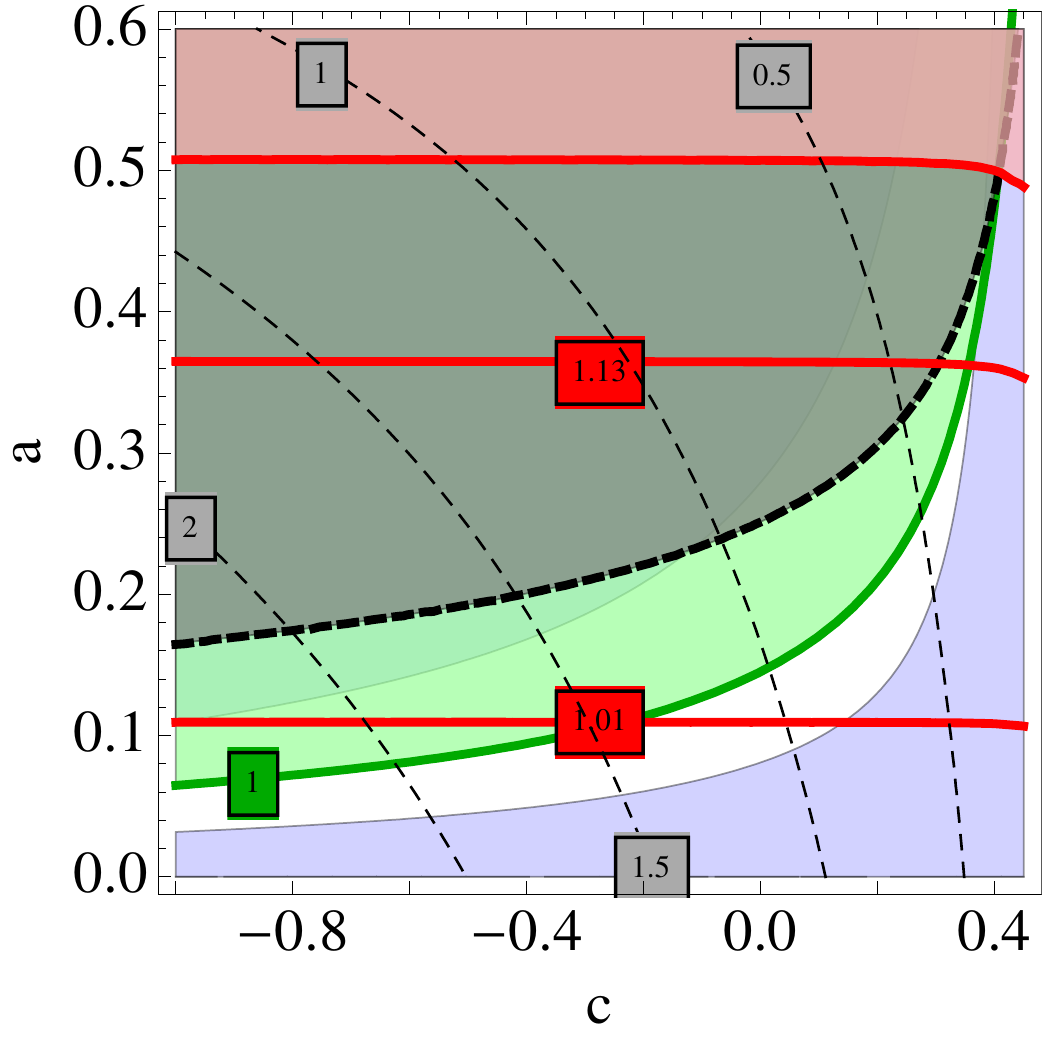} \hfill
\caption{\it Contour plot of $\Delta a_\mu$ along with bounds from oblique corrections (dashed black line) stability bound (solid green line) and from the Higgs strength into diphotons (solid red upper bound). We have considered  $c_{\mu_L}=0.4$, $c_{\mu_R}=0.5$.}
\label{fig:final}
\end{figure} 

One possible solution to accommodate the LHCb anomaly requires the presence of massive vector gauge bosons which are strongly coupled to muons and very weakly coupled to electrons, thus breaking lepton universality. In the considered theory the massive vector gauge bosons are naturally identified with the KK modes of the $Z$ and photon gauge bosons, whose couplings with fermions depend on their degree of (compositeness) IR localization: the more composite the fermions the more strongly coupled they are to KK modes. Thus a simple and natural solution to the LHCb anomaly is considering muons more composite than electrons. This solution has been proven to be consistent with all electroweak constraints by a simple choice of the localizing parameters for the muon and the bottom quark $c_{\mu_{L,R}},c_{b_{L,R}}$. In particular the adopted values in this paper are
\begin{align}
c_{\mu_L}&=0.4,\ c_{\mu_R}=0.5\nonumber\\
c_{b_L}&=0.44,\ c_{b_R}=0.58\nonumber
\end{align}

Although this minimal theory has all the ingredients to also solve the muon AMM problem, it fails to provide a strong enough chirality flip to cope with the experimental value of $a_\mu$. In order to do that we have enlarged the theory with a set of vector-like leptons, a doublet and a singlet, mixed with the muon sector through Yukawa interactions. The VLL propagate in the bulk with localizing parameters $c_L$ and $c_R$ for the doublet and singlet, respectively. The use of VLL (unmixed with the muon sector) has been often proposed in the past to increase the value of the width $H\to\gamma\gamma$ in order to cope with a possible deviation with respect to the SM prediction~\cite{ArkaniHamed:2012kq,Joglekar:2012vc}. In our case VLL are mixed with the muon sector, which implies strong constraints, not only from universal (oblique) observables but also from non-oblique ones, in particular from the $Z\overline\mu\mu$ coupling. This exercise has been performed in this paper where we show that a region in the space of parameters ($c_L,c_R$) is consistent with the muon AMM value and all experimental and theoretical constraints. The original region consistent with the muon AMM is in fact restricted by all electroweak constraints. 
For the particularly simple case of equal $c_L=c_R\equiv c$ the combined allowed region is given by the plot in Fig.~\ref{fig:final},
where we have superimposed the region allowed by electroweak precision observables (dashed black line) as well as the region allowed by the Higgs strength into diphotons (upper red solid line) and the region allowed by the stability of the electroweak vacuum (solid green line). As we can see the present bound from $H\to\gamma\gamma$ is superseded by the other constraints. However if in the future the Higgs strength approaches the SM value ($\widehat\mu=1$) it can become the strongest constraint. One general consideration from Fig.~\ref{fig:final} is that the available region implies:
\begin{itemize}
\item
That VLL are localized toward the IR brane. In the dual theory it means that VLL have a high degree of compositeness. In particular $c\lesssim 0.37$.
\item
There is a lower bound on the mass of VLL. In particular the lightest VLL mass is $M_{\widetilde{\mathcal L}}\gtrsim 270$~GeV.
\end{itemize} 
 
 A smoking gun for this theory would be, apart from the direct detection of a KK mode at $\sim 2$ TeV, the direct detection of a charged or neutral VLL. In fact VLL are produced at the LHC by Drell-Yan production and their present bound at 95\% CL, based on the ATLAS analysis at $\sqrt{s}=8$~TeV and an integrated luminosity of 20.3 fb$^{-1}$, is the mild one $M_{\widetilde{\mathcal L}}\gtrsim 168$~GeV. However with increasing luminosity and center of mass energy  $\sqrt{s}=13$~TeV we expect the bounds will rapidly improve.
 
The last point we want to comment is the capability of this theory to encompass dark matter (DM) with the cosmological abundance consistent by WMAP results $h^2 \Omega\simeq 0.12$~\cite{Agashe:2016kda}. In our theory (with 5D SM plus VLL) there is no candidate to DM as the lightest VLL, $\widetilde{\mathcal L}$, decays with a width $\propto (U_L^{31}U_R^{21},|U_L^{31}|^2+|U_R^{21}|^2)$. One could then enlarge the theory with a new VLL$'$ sector (where a discrete symmetry prevents the mixing with the SM leptons) which includes a (sterile) singlet $S'(x,y)\propto \mathcal S'(x)$ (the 5D counter-part of the 4D right-handed neutrino field), as in Ref.~\cite{vonGersdorff:2012tt}, mixed with the VLL$'$ active neutrino $N'$ by the 5D Lagrangian
\be
\mathcal L_5= \widehat Y_S\overline D'(x,y) \sigma H^\dagger(x,y) S'(x,y)
\ee
and arrange that the lightest VLL$'$ be a linear combination of the field $\mathcal N'$ (member of an $SU(2)_L$ doublet) and the 4D component of the sterile neutrino (along the lines of Ref.~\cite{Joglekar:2012vc} in 4D theories). Direct searches exclude DM which is mostly $\mathcal N'$, as it has unsuppressed couplings with the $Z$ and thus large interaction rates with nucleons, but states which are mostly $\mathcal S'$ provide very small annihilation rates, and then lead to large relic densities which rapidly overclose the universe, unless annihilation is enhanced by resonant and/or co-annihilation effects. A thorough analysis of DM in our theory is beyond the scope of this paper and will be postponed for future investigation.

\section*{\sc Acknowledgments}

We thank M. Beneke for comments, and especially G.~Panico and
O.~Pujol\`as for many useful discussions and for having participated
in the early stages of this work.  L.S. is supported by a Beca
Predoctoral Severo Ochoa del Ministerio de Econom\'{\i}a y
Competitividad (SVP-2014-068850), and E.M. is supported by the
European Union under a Marie Curie Intra-European fellowship
(FP7-PEOPLE-2013-IEF) with project number PIEF-GA-2013-623006, and by
the Universidad del Pa\'{\i}s Vasco UPV/EHU, Bilbao, Spain, as a
Visiting Professor. The work of M.Q. and L.S.~is also partly supported
by Spanish MINECO under Grant CICYT-FEDER-FPA2014-55613-P, by the
Severo Ochoa Excellence Program of MINECO under the grant
SO-2012-0234, by Secretaria d'Universitats i Recerca del Departament
d'Economia i Coneixement de la Generalitat de Catalunya under Grant
2014 SGR 1450, and by the CERCA Program/Generalitat de Catalunya. The
research of E.M. is also partly supported by Spanish MINECO under
Grant FPA2015-64041-C2-1-P, by the Basque Government under Grant
IT979-16, and by the Spanish Consolider Ingenio 2010 Programme CPAN
(CSD2007-00042).


\end{document}